\documentclass[12pt]{article}

\pdfoutput=1

\usepackage{graphicx}
\usepackage{amssymb}
\usepackage{amsmath}

\usepackage{braket}

\usepackage{bm}

\newcommand{\rf}[1]{(\ref{#1})}
\newcommand{\beq}{\begin{equation}}

\newcommand{\eeq}{\end{equation}}
\newcommand{\bea}{\begin{eqnarray}}
\newcommand{\eea}{\end{eqnarray}}

\newcommand{\n}{\nu}
\newcommand{\m}{\mu}

\newcommand{\Del}{\Delta}

\newcommand{\tr}{\mathrm{tr}\,}
\newcommand{\ra}{\rangle}
\newcommand{\la}{\langle}

\newcommand{\cT}{{\cal T}}

\newcommand{\cN}{{\cal N}}

\begin{document}

\begin{center}
\vspace{24pt}
{ \large \bf The effective action in  4-dim CDT. 

The transfer matrix approach.  }

\vspace{30pt}

{\sl J. Ambj\o rn}$\,^{a,b}$,
{\sl J. Gizbert-Studnicki}$\,^{c}$,
{\sl A. G\"{o}rlich}$\,^{a,c}$,
{\sl J. Jurkiewicz}$\,^{c}$

\vspace{24pt}
{\footnotesize

$^a$~The Niels Bohr Institute, Copenhagen University\\
Blegdamsvej 17, DK-2100 Copenhagen \O , Denmark.\\
{ email: ambjorn@nbi.dk}\\

\vspace{10pt}

$^b$~IMAPP, Radboud University\\
 Niemegen, The Netherlands, \\

\vspace{10pt}

$^c$~Institute of Physics, Jagellonian University,\\
Reymonta 4, PL 30-059 Krakow, Poland.\\
{ email: jakub.gizbert-studnicki@uj.edu.pl, atg@th.if.uj.edu.pl, jurkiewicz@th.if.uj.edu.pl}\\

\vspace{10pt}

}
\vspace{48pt}

\end{center}


\begin{center}
{\bf Abstract}
\end{center}

We measure the effective action in all three phases 
of 4-dimensional Causal Dynamical Triangulations (CDT) 
using the transfer matrix method. The transfer matrix is  
parametrized by the total 3-volume of the CDT universe at a 
given (discrete) time. We present a simple effective model based 
on the transfer matrix measured in the de Sitter phase. It allows us 
to reconstruct the results of full CDT in this phase.  
We argue that the transfer matrix method is valid not only inside the  
de Sitter  phase ('C') but also in the other two phases.  
A parametrization of the measured transfer matrix/effective action 
in the  'A' and 'B' phases is proposed and the  relation 
to phase transitions is explained. We discover a potentially 
new 'bifurcation' phase separating the  de Sitter phase ('C') and  
the 'collapsed' phase ('B').

\section{Introduction}

The method of triangulations was introduced in the context of  General Relativity by Regge \cite{regge} to discretize  the continuous Hilbert-Einstein action 
\begin{equation}
S_{HE}[g]=\frac{1}{16 \pi G}\int{d^4x \sqrt{-g}(R-2 \Lambda)}. 
\label{SHE}
\end{equation}
Continuous geometries are approximated by piecewise linear  
simplicial manifolds.  The curvature is represented as a 
deficit angle concentrated at the (D-2) subsimplex. 
A particular realization of this idea is the method of Dynamical
Triangulations (DT), where the piecewise linear simplicial manifolds 
are built by gluing together regular, identical simplices with 
identical edge lengths $a$ \cite{ET2d}.
The DT set of simplicial geometries is thus entirely characterized by 
the abstract triangulations which define how the simplices are glued together and it has been useful in Monte Carlo simulations of quantum gravity, $a$ acting as a UV cut off. 
In D=2 one could even solve the DT model analytically for gravity coupled to certain simple 
matter systems and the continuum limit $a \to 0$ could be 
obtained. These results were reproduced by conformal field theory
methods (so-called 2D quantum Liouville theory) \cite{ET2Dmatter}.  

For higher dimensional quantum gravity the DT approach 
has been less successful \cite{ET3D,ET4D}. Firstly, there are only 
very few analytical results. Most investigations use 
Monte Carlo simulations to evaluate the path integrals.
This method has also been tested and has proven very successful 
in $D=2$. Secondly, in the three-- and four--dimensional DT cases  
the simplest versions of the lattice theory, characterized
by two coupling constants, analogous to that of the continuum theory  
(\ref{SHE}), did not show a behaviour which could be viewed as
interesting from a continuum gravity point of view.
Depending on the strength of the bare lattice gravitational 
coupling constant, the system (rotated to Euclidean time in order
to allow for Monte Carlo simulations) appeared to have two phases. 
The weak gravity phase was dominated by the branched polymer geometries 
with a Hausdorff dimension  $d_H = 2$
and the strong gravity phase by collapsed geometries with possibly $d_H = \infty$, 
corresponding to universes without a linear extension. The two phases
were separated by a first order phase transition \cite{ET1stOrder}.

The method of Causal Dynamical Triangulations (CDT) was introduced to cure 
these problems \cite{CDT} (for pedagogical reviews see \cite{CDTreviews}).
At this point it should be made clear that the problems encountered 
in DT could very well reflect the fact that there {is no}
stand alone theory of quantum gravity based only  on the  metric tensor 
$g_{\m\n}$. This is in a certain way what we are trying to investigate.
CDT enlarges the scope of metric theories one can reach, but eventually
one might encounter some of the same problems as in DT. 
In CDT a notion of the proper time was introduced together with the 
requirement that the spatial topology of the quantum universe with 
respect to this proper time 
must be preserved in the time evolution \cite{causality}. 
The simplest version of the discretized theory using the CDT 
approach has three parameters. Apart from the two parameters
present in the DT approach, related to the cosmological constant and 
to the gravitational constant, the additional parameter controls 
a possible asymmetry between the
edge lengths in the spatial and time directions. 
In numerical simulations the topology of the manifold is assumed 
to be $S_3 \times S_1$ with periodic boundary conditions in the 
(Euclidean) time. This choice
is dictated by practical reasons. Geometric structures 
used to build simplicial manifolds of CDT are characterized by their 
position in spatial and time directions.
In particular we use two types of  four-simplices: $\{4,1\}$-simplices with 
four vertices at time $t$ and one at $t\pm 1$ and $\{3,2\}$-simplices with 
three vertices at $t$ and two
at $t\pm 1$. All simplices of a particular type are assumed to have 
the same sizes. The discretized Regge action in this case takes a 
form \cite{CDTreviews}:
\begin{equation}
S_R =  -(K_0+6\Delta) N_0 + K_4 \left( N^{\{4,1\}}+ N^{\{3,2\}}\right)+\Delta  N^{\{4,1\}}
\label{Sdisc}
\end{equation}
where  $N_0$ is the total number of vertices in the triangulation,  
$N^{\{4,1\}}$ and $ N^{\{3,2\}}$ are the total numbers of simplices of 
type $\{4,1\}$ and type $\{3,2\}$, respectively. $K_0$, $K_4$ and $\Delta$ 
are the (bare) dimensionless coupling constants obtained by the 
discretization of the continuous action (\ref{SHE}). An additional 
geometric parameter is the length $t_{tot}$ of the periodic time axis.

In numerical simulations the total four-volume of the universe is kept fixed. 
In practice this restricts the number $N^{\{4,1\}}$ to fluctuate around a 
fixed value $ \bar{N}_{41}$. For a fixed
space-time topology the number of CDT triangulations  
with  $ \bar{N}_{41}$ $\{4,1\}$ simplices  grows exponentially 
with $ \bar{N}_{41}$. This exponential growth determines the critical value 
$K_4^{crit}$ of the bare lattice cosmological constant $K_4$.   
Requiring the average $\langle  N^{\{4,1\}} \rangle$ to be fixed
is equivalent to fixing the bare cosmological constant $K_4$ 
to be close to the critical value $K_4^{crit}$. 
For $K_4 < K_4^{crit}$ the {\em partition function} 
\begin{equation}
{\cal Z} = \sum_{{\cal T}} e^{-S_R}
\end{equation}
becomes divergent.

\begin{figure}[h!]
\centering
\scalebox{0.7}{\includegraphics{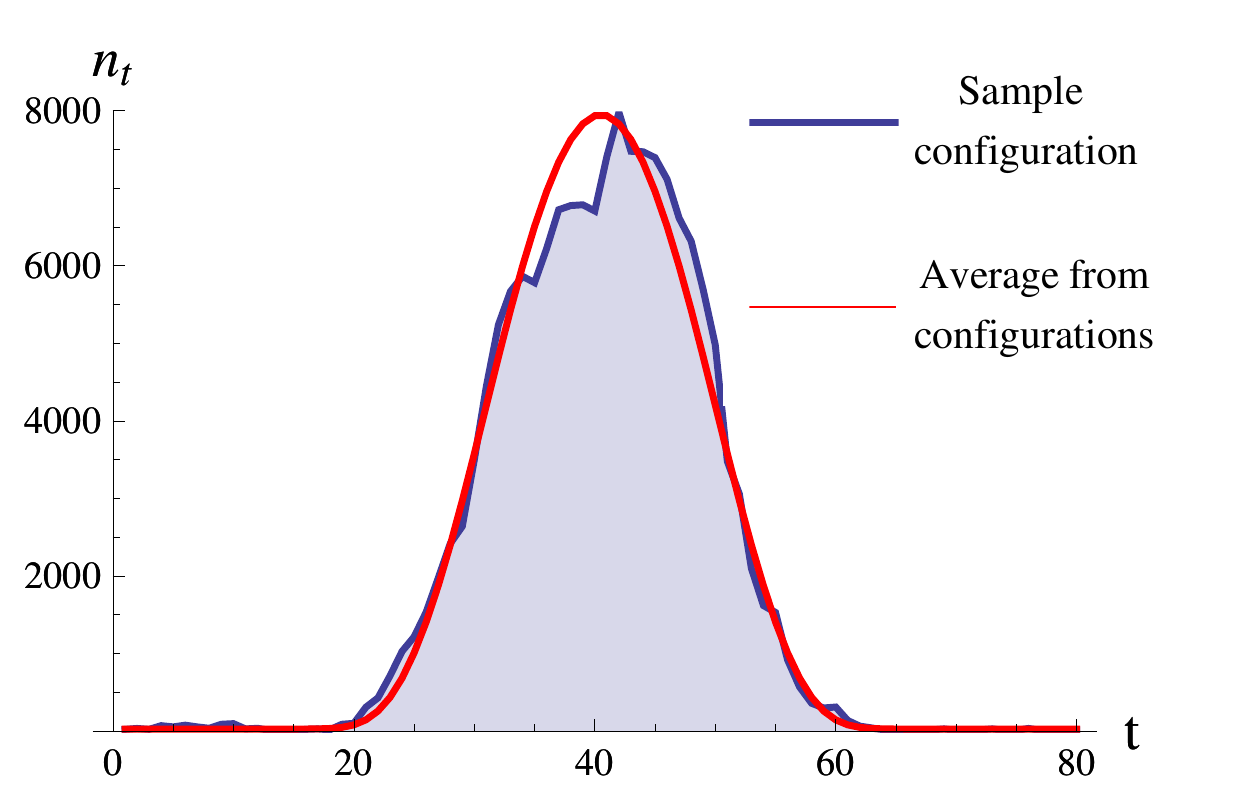}}
\caption{The measured distribution $n_t$ of a \{4,1\} 
volume inside the de Sitter phase. The data are 
for coupling constants $K_0 = 2.2$ and $\Delta =0.6$. 
The blue line represents a 
single configuration generated in  the Monte Carlo simulations. 
The red line represents the 
distribution, averaged over many configurations in a way described in
\cite{bigs4}.}
\label{Fig01}
\end{figure}

The remaining two bare coupling constants can be chosen freely 
and the choice will determine the ``physics'' of  the (lattice) theory. 
Numerical simulations proved 
that depending on these values the  quantum system can be in three different
phases \cite{CDTphases}. From a physical point of view the most interesting 
is the {\em de Sitter phase}, where a typical geometry can be viewed 
as a quantum fluctuation around a semi-classical regular
four-sphere, or rather four-ellipsoid with different scale in time 
and spatial directions \cite{deSitter}. 
A natural quantity used to parametrize the configurations is the 
distribution $n_t$ of  $\{4,1\}$ volumes as a function of the 
discrete time $t$. 
$n_t$ is closely related to the scale factor in the minisuperspace approach to
quantum gravity. $n_t$ is  equal to twice the 
number of three-dimensional tetrahedra which form a  
spatial slice (with $S^3$ topology) at time $t$. This is because 
each spatial tetrahedron located in time-slice $t$ is shared by 
precisely two $\{4,1\}$ four-simplices, one with its fifth  vertex 
at time-slice $t+1$
and one with its fifth vertex at time-slice $t-1$ (and of course both 
four-simplices have four vertices 
at time-slice $t$, namely the four vertices of the spatial tetrahedron
they share).   
In the de Sitter phase the distribution of $n_t$
has a characteristic shape  (Fig. \ref{Fig01}), 
consisting of the {\em blob} which fluctuates around a four-sphere 
and a {\em stalk} with an almost minimal size. The stalk is present
because the global topology
$S_3 \times S_1$ is not allowed to be broken in the computer simulations. 
The distribution of spatial volume can be averaged over many 
statistical  independent configurations obtained in the numerical simulations. 
In the blob the average spatial volume profile 
$\braket{n_t}\propto \cos^3(t/B)$ which corresponds to a (Wick rotated) 
de Sitter solution of Einstein's equations. 
Fluctuations around this semi-classical trajectory 
 $\Delta n_t = n_t - \langle n_t\rangle$
are correlated for different $t$. The covariance matrix of fluctuations can be measured. 
The inverse of the covariance matrix can be used to determine 
the {\em effective action} in terms of $n_t$ \cite{bigs4,CDTreviews}. 
It was shown in \cite{we}
that it corresponds to a naively discretized minisuperspace action
\begin{equation}
S_{eff}^{blob}=\frac{1}{\Gamma}\sum_t \left( \frac{\left( n_t-n_{t+1}\right)^2}{n_t + n_{t+1}} - \lambda \, n_t + \mu \, n_t^{1/3} + {\cal O}(n_t^{-1/3}) \right) \ ,
\label{Seff}
\end{equation}
where $\Gamma$ is proportional to the effective 
Newton's constant while the effective cosmological 
constant $\lambda$ together with the parameter $\mu$ 
fix the total 4-volume of the universe.

This  form of the action leads in a natural way 
to a path-integral representation with the weight $\exp(-S_{eff})$ 
of each configuration given by  a product
\begin{equation}
\exp(-S_{eff}) \equiv \prod_t \exp(-L_{eff}(n_t,n_{t+1}))
\label{TMS}
\end{equation}
of pseudo-local transfer matrix elements 
\begin{equation}
\bra{n_t}  M \ket{n_{t+1}}\propto  \exp(-L_{eff}(n_t,n_{t+1}))
\label{TMLag}
\end{equation}
linking neighbouring spatial slices. In this form 
all details of the geometric structure space at a given 
spatial slice are wiped out and we assume that it makes sense to
use the {\em effective} quantum states $| n_t \rangle$ 
with unit norm as an eigenstate basis at each slice.

The transfer matrix can be measured in numerical simulations. 
In \cite{TM1} we used this concept to determine the  form of the 
effective action inside the de Sitter phase. We found that the 
symmetrized  form of (\ref{Seff}) with minor small-volume correction
fits very well to our numerical data both in the blob and the stalk range 
of the CDT universe. This result was obtained in numerical simulations of 
systems with small time extension ($t_{tot}=2,3,4$). 

In CDT there 
exists a  ``genuine'' transfer matrix $M_{gen}$ connecting states
at time $t$ and time $t+1$. These states can be chosen as the states
of spatial geometries, and in this 
approach a given spatial  
geometry is completely characterized be  the corresponding 
$DT$ triangulation of $S^3$ (which is part of the 4d CDT triangulation). 
Thus we have by definition for the ``genuine'' transfer matrix:
\begin{equation}\label{jan1}
{\la T(t+n)| M_{gen}^n|T(t)\ra =}
\end{equation}
$$
=\sum_{T(t+i), 1<i<n-1} 
\la T(t+n)| M_{gen}|T(t+(n-1))\ra\cdots \la T(t+1)| M_{gen}|T(t)\ra \ .
$$

The number of states $|T(t)\ra$ is of course much larger than the 
number of so-called ``effective'' quantum states $|n_t\ra$ mentioned
above. The claim  that the effective transfer matrix describes 
CDT well contains two aspects, namely
\beq\label{jan2}
\la T(t+1)| M_{gen}|T(t)\ra \sim \la n_{t+1}| M |n_t\ra,
\eeq
for generic states $|T\ra$, as long as we only measure
$n_t$, and even stronger 
\beq\label{jan3}
\la T(t+n)| M_{gen}^n|T(t)\ra \sim  \la n_{t+n}| M^n |n_t\ra,
\eeq  
again when we only look at $n_t$.
To be reassured 
that the effective transfer matrix approach is correct we have 
to check \rf{jan3} for large $n$, that is for systems where 
$t_{tot}\gg 1$. The aim is to reproduce the full CDT results 
(the volume profile $\braket{n_t}$  and quantum fluctuations 
$\Delta n_t$) by studying  the simplified effective model 
based on the measured transfer matrix.

We also want to extend our analysis of the 
effective transfer matrix/action to two other phases of CDT  
which are the analogues of the branched polymer and collapsed phases of DT. 
This is especially interesting in the context of phase transitions. 
Recent results \cite{phases1} showed that the phase transition 
between the de Sitter phase and the collapsed phase is a 
second (or higher) order phase transition. 
This makes it a natural candidate in the quest for UV fixed points of CDT. 
Therefore it is important to understand better  the nature 
of  the CDT phase transitions from a microscopic perspective.

\section{Methodology of the transfer matrix measurements}


In order to investigate the properties of CDT  in four dimensions 
we have performed computer simulations of systems with (time) periodic 
boundary conditions and $S^3$ spatial topology. 
The action used in the computer simulations is the Regge discretization 
of the Einstein-Hilbert action, given by Eq.\  (\ref{Sdisc}).

Studies of the covariance matrix of spatial volume fluctuations, 
$$
C_{t t'} \equiv \langle (n_t - \langle n_t \rangle) 
(n_{t'} - \langle n_{t'} \rangle) \rangle,
$$
suggest that the effective action couples only adjacent time slices 
and that there  exists an effective transfer matrix, namely the one  
defined  by Eq.\ (\ref{TMLag}).

Inside the de Sitter phase (also called phase 'C')  
the measurement and para-metrization of the 
transfer matrix is  straightforward.  In the other two phases it has to 
be done with some care. The most problematic phase is the 'time collapsed' 
phase (also called phase 'B'), in which  time translation symmetry is 
strongly broken in generic triangulations. 
Measurements inside phase 'B'  required modifying 
the Monte Carlo code used in earlier computer simulations. 
The new measurement method uses a system with just two time slices, 
and one has to avoid artificial repetition of (sub)simplices 
(the problem does not occur  when the  time direction has more
than  two time slices). 
We checked that inside the de Sitter phase the results of 
the new method are fully consistent with previous results 
based on the systems with 3 or  4 time slices.


In order to measure the transfer matrix we need systems with a small 
time period $t_{tot}$. In our transfer matrix parametrization 
the probability to measure the combination of 3-volumes 
$n_t=N^{\{4,1\}}(t)$ in times $t=1\dots t_{tot}$ is given by: 
\begin{equation}
{P}^{(t_{tot})} (n_1,n_2,...,n_{t_{tot}} ) = \frac{\braket{n_1|M|n_2}\braket{n_2|M|n_3}...\braket{n_{t_{tot}}|M|n_1}}{ \tr M^{t_{tot}} }\ .
\label{Pn1nttot}
\end{equation}
In a system with  two time slices ($t_{tot}=2$) we have:
$$
{ P}^{(2)} (n_1,n_2) = \frac{\braket{n_1|M|n_2}\braket{n_2|M|n_1} }{ \tr M^{2} } \ ,
$$
which implies:
\begin{equation}
\braket{n|M|m} \propto {\sqrt{P^{(2)} (n_1=n,n_2=m ) }} \ ,
\label{M2}
\end{equation}
where we use the assumption that due to 
time-reflection symmetry the transfer matrix is also symmetric.

We can also use the probability distributions measured in Monte Carlo 
simulations with $t_{tot}=3$ and $4$ (this was done in our earlier 
investigations). In this approach the transfer matrix elements 
can be computed as 
\begin{equation}
\braket{n|M|m}\propto \frac{P^{(3)} (n_1=n,n_2=m ) }{\sqrt{P^{(4)} (n_1=n,n_3=m ) }} \ .
\label{M34}
\end{equation} 
We checked that both approaches agree inside the de Sitter phase 'C'.

The advantage of the new method with $t_{tot}=2$ is twofold. 
First of all one needs only to  measure a single probability distribution, 
thus leading to a reduction of computer time and to smaller  
statistical errors 
(since one does not  need to combine two probability measurements like in 
\rf{M34}). 
However,  more importantly, by an appropriate choice of  volume 
fixing (see below) one can measure off-diagonal elements of the 
transfer matrix with much higher precision. 
It is especially important when extracting the kinetic part of 
the effective action in   the  'B' phase and in the third phase
(which is called the 'A' phase).
 
To perform the computer simulations efficiently one  
has to  introduce some kind of volume fixing. This is done 
by adding to  the usual Regge action (\ref{Sdisc}) an 
additional volume fixing term:
$$
S_R \to S_R + S_{VF} \ .
$$ 
In our simulations with $t_{tot}=2$  we use the global volume 
fixing\footnote{Our previous approach used (\ref{M34}) based on probability 
distributions measured in systems with $t_{tot}=3,4$, and we 
used a {\it local} volume 
fixing procedure (see \cite{TM1} for details). The transfer matrix measurement 
with global volume fixing is possible only with $t_{tot}=2$ and is 
especially suitable in the 'A' and 'B' phases  where generic configurations 
typically have very  different spatial volumes in neighbouring 
time slices.} either with a quadratic or a linear potential:
\begin{equation}
S_{VF}= \epsilon (n_1+n_2-n_{vol})^2 \quad \textrm{or} \quad S_{VF}= \epsilon |n_1+n_2-n_{vol}| \ .
\label{VFpot1}
\end{equation}
The effect of the volume fixing term can be easily removed from 
the measured transfer matrix $\widetilde M$ defined  by (\ref{M2}) by setting:
\begin{equation}
\braket{n|M|m}= e^{\frac{1}{2} \epsilon (n+m-n_{vol})^2 }  
\braket{n|\widetilde M|m}  \ \textrm{or} \ 
\braket{n|M|m}= e^{\frac{1}{2} \epsilon |n+m-n_{vol}| }  \braket{n|\widetilde M|m} \ ,
\label{Mcor}
\end{equation}
for a quadratic or a linear volume fixing, respectively.

The volume fixing correction (\ref{Mcor}) clearly affects  
the diagonal elements of the transfer matrix used in the analysis 
of the the potential term (see below), whereas the cross-diagonal 
elements, important for the  determination of  the kinetic 
term,  are  simply rescaled  for $n+m=const$.


\section{The effective model in the de Sitter phase}

\begin{figure}
\begin{center}
\includegraphics[width=0.7\textwidth]{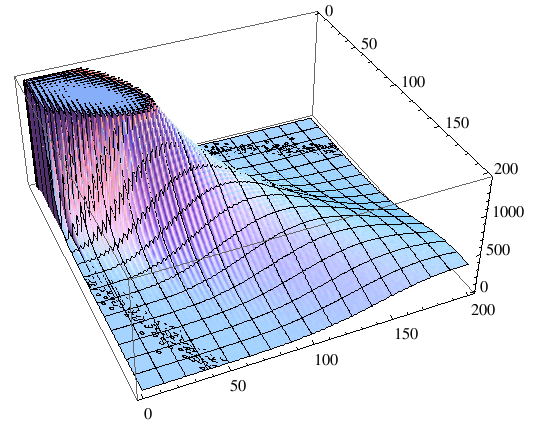}
\end{center}
\caption{Measured transfer matrix elements $\langle n | M | m \rangle$ in the de Sitter phase 'C'.
For small values of $n$ and $m$ we observe strong discretization effects.
For larger volumes the behaviour is smooth.}
\label{fig:transfer}
\end{figure}

Recent results show that for small $t_{tot}$
the measured transfer matrix in the de Sitter phase does not depend 
on the number of slices
supporting the decomposition (\ref{Pn1nttot}).
An example of the measured transfer matrix is plotted in Fig. \ref{fig:transfer}.

\begin{figure}
\begin{center}
\includegraphics[width=0.7\textwidth]{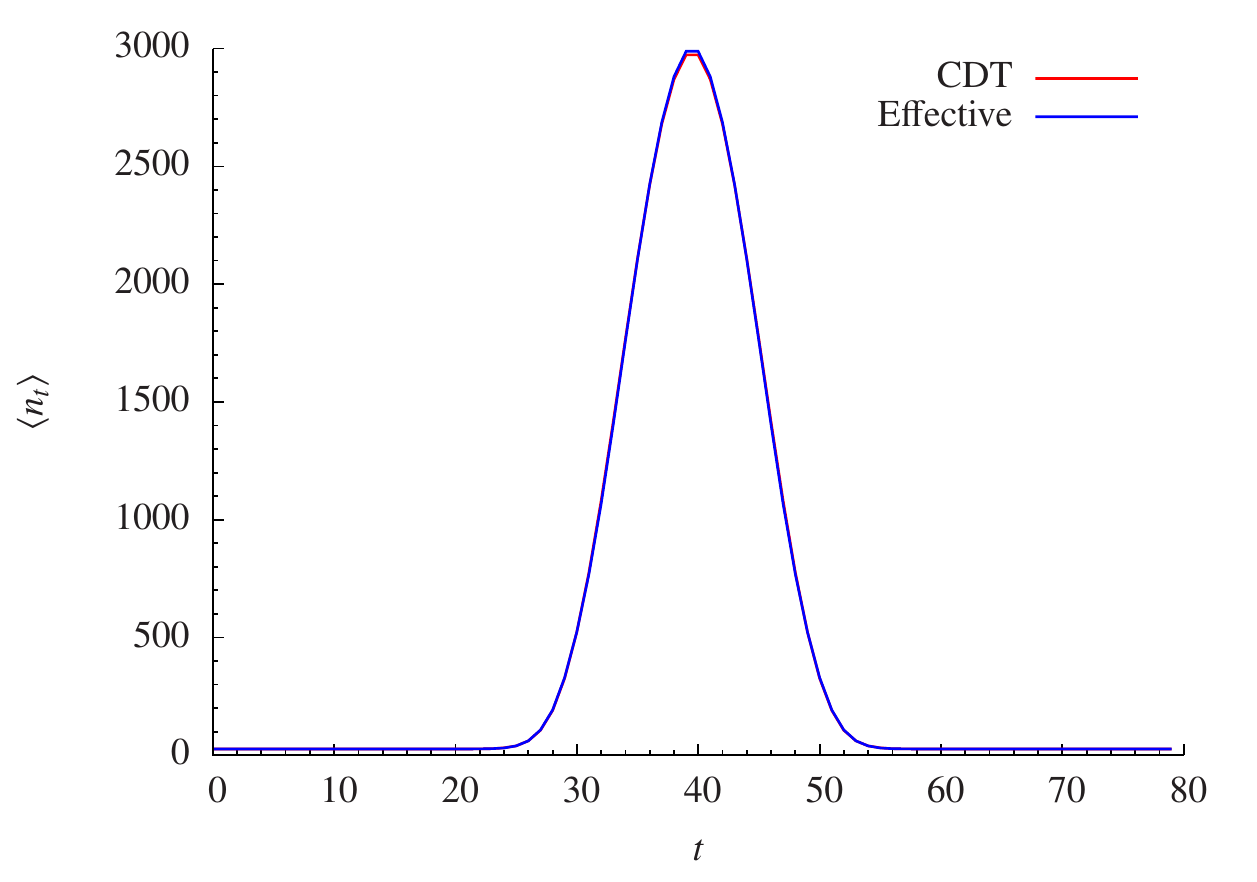}
\end{center}
\caption{The average volume profile $\langle n_t \rangle$ for full CDT in the de Sitter phase (red) and the effective model (blue) for $t_{tot}=80$ and $\bar{N}_{41}=40$k.}
\label{fig:avt}
\end{figure}

In order to reconstruct the results of the full CDT simulations 
we will need matrix 
elements $M_{nm}\equiv\langle n | M | m \rangle$ for large volumes $n, m$.
As can be seen from Fig. \ref{fig:avt}, for a total 
four-volume with $\bar{N}_{41}= 40\mathrm{k}$ the 
largest spatial volumes $n_t$ reach values above $3000$.
Technically it is difficult to measure matrix elements in 
such wide range, but we can use extrapolation for large volumes.

For small volumes the transfer matrix elements 
are dominated by very strong discretization effects 
as can be seen  in Fig. \ref{fig:transfer},
 but as $n$ and $m$ increase the behaviour becomes much smoother.
For sufficiently large spatial volumes the transfer matrix 
is very well described by the effective Lagrangian introduced in \cite{TM1},
\begin{equation}
L_{eff}(n,m)=
\label{Seffform}
\end{equation}
$$ 
=\frac{1}{\Gamma} \left[ 
\frac{(n - m)^2}{n + m - 2 n_0} + \mu \left( \frac{n + m}{2}\right)^{1/3} - \lambda \left( \frac{n + m}{2}\right) - \delta \left( \frac{n + m}{2}\right)^{- \rho}  \right] .
$$
We use the measured (empirical) transfer 
matrix $M^{(\mathrm{emp})}_{n m}$, $250 < n, m < 700$, to determine 
the parameters $\Gamma, n_0, \mu, \lambda, \delta$ and $\rho$, by making 
a best fit of 
\begin{equation}
 M^{(\mathrm{th})}_{n m} = \cN e^{- L_{eff}(n, m)} \ ,
 \label{eq:Mth}
\end{equation}
to $M^{(\mathrm{emp})}_{n m}$. For larger values of $n,m$ we then 
use $M^{(\mathrm{th})}_{n m}$, with $L_{eff}(n,m)$ determined by  
this fit.

Finally, we thus define the semi-empirical transfer matrix  by
\begin{equation}
M_{n m} = \left\{ \begin{array}{ll}
M^{(\mathrm{emp})}_{n m} & n < thr \textrm{ or } m < thr,\\
M^{(\mathrm{th})}_{n m} & \textrm{otherwise},
\end{array} \right.
\label{Mextr}
\end{equation}
where $thr$ is a threshold ($thr = 300$).
When one of the entries is smaller than the threshold we 
use the  measured matrix elements.
When both entries are larger than the threshold we 
use the extrapolating function (\ref{eq:Mth}).
\begin{figure}
\begin{center}
\includegraphics[width=0.49\textwidth]{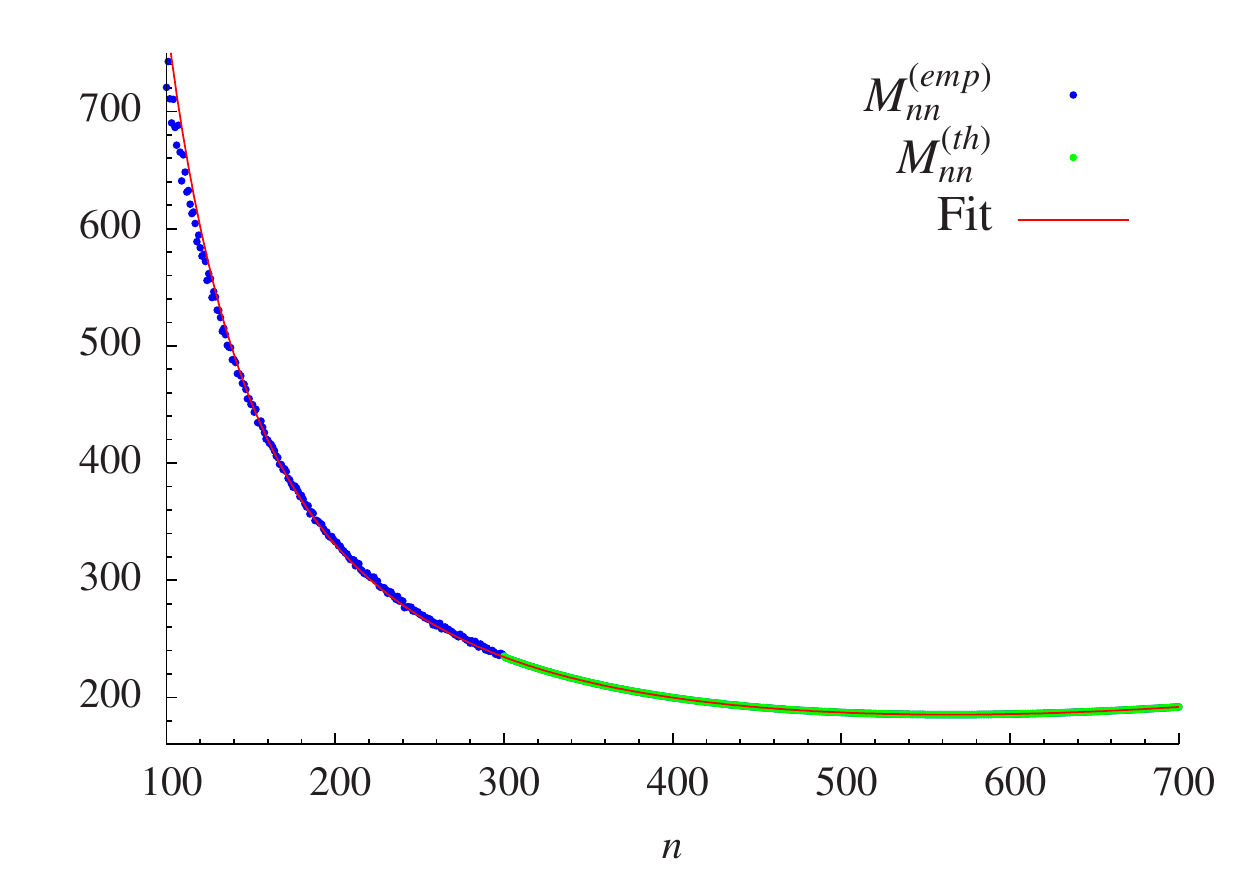}
\includegraphics[width=0.49\textwidth]{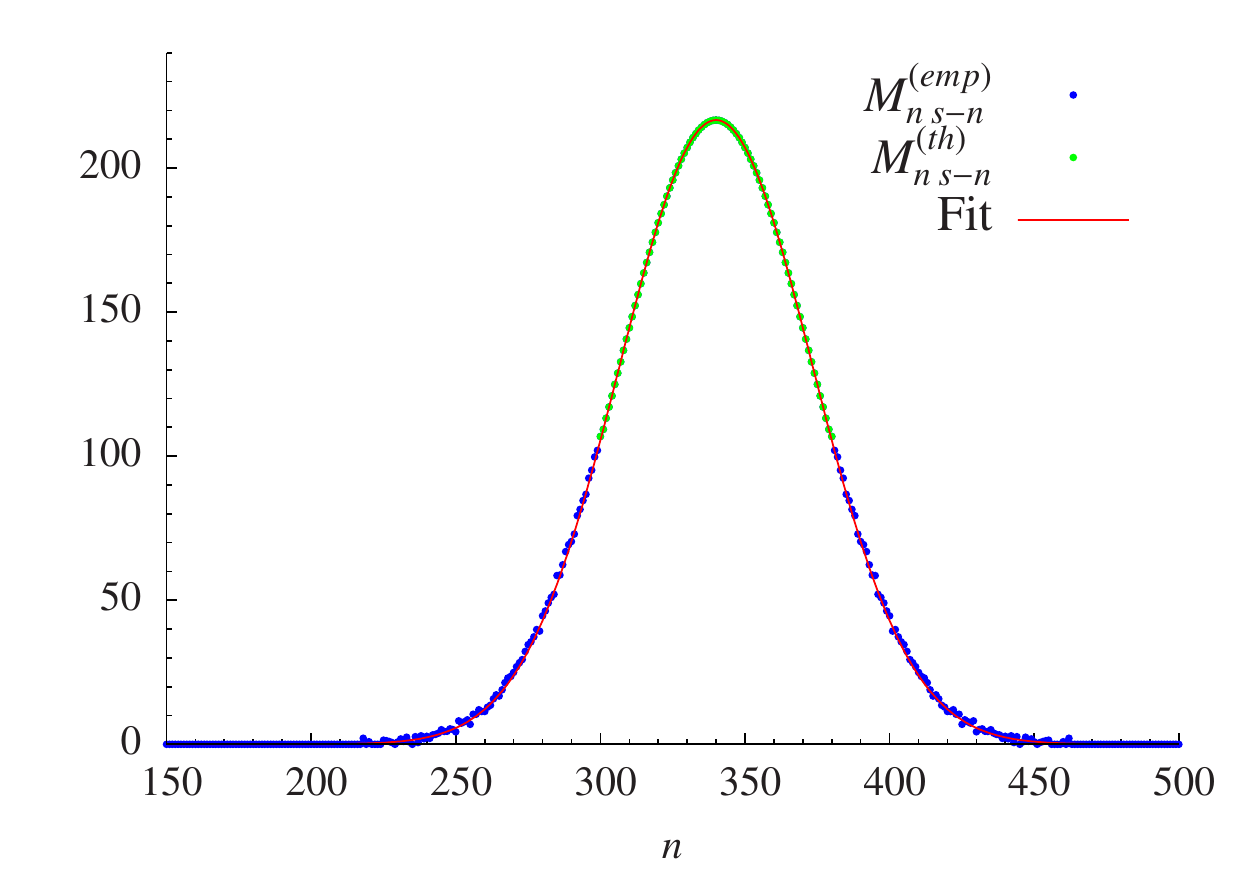}
\end{center}
\caption{
The effective transfer matrix in the 
de Sitter phase, merged from the empirical matrix (blue) 
and theoretical matrix (green). The theoretical matrix (red line) is 
determined by a best fit to the empirical matrix in an overlap 
region where $n$ and $m$ are in the range 250-700 as described in the text.
Left: Diagonal $\langle n | M | n \rangle$.
Right: Anti-diagonal $\langle n | M | s- n \rangle$, $s = 680$.
}
\label{fig:semi}
\end{figure}
Fig. \ref{fig:semi} presents the diagonal $\langle n | M | n \rangle$ (left) 
and an anti-diagonal $\langle n | M | s - n \rangle $ (right) 
of the semi-empirical transfer matrix.
The blue points denote the empirical part of $M$, used 
for volumes below the threshold, i.e.\ for $n < 300$ or $m < 300$.
The red line presents a theoretical fit of the form (\ref{eq:Mth}), 
the fitting range being $250 - 700$.
The green points correspond to 
the theoretical part of $M$ given by (\ref{eq:Mth})
for volumes above the threshold, i.e.\ for $n$ and $m$ larger than  300.
The agreement between the empirical transfer matrix and 
the fit (plotted as the  red line)
is very good also below $n = 250$ but gets even better above the threshold.
The extrapolation (\ref{eq:Mth}) allows us 
to expand the transfer matrix to volumes
which are not easily accessible by direct measurement.
Nevertheless, because the effective Lagrangian 
describes perfectly the measured transfer matrix in the range where 
we can make the comparison, and seemingly gets better 
with increasing  values of the entries $(n,m)$, this extrapolation beyond
actual empirical data should not be of 
any importance when judging the validity of the 
effective transfer matrix decomposition (\ref{Pn1nttot}).

\begin{figure}
\begin{center}
\includegraphics[width=0.7\textwidth]{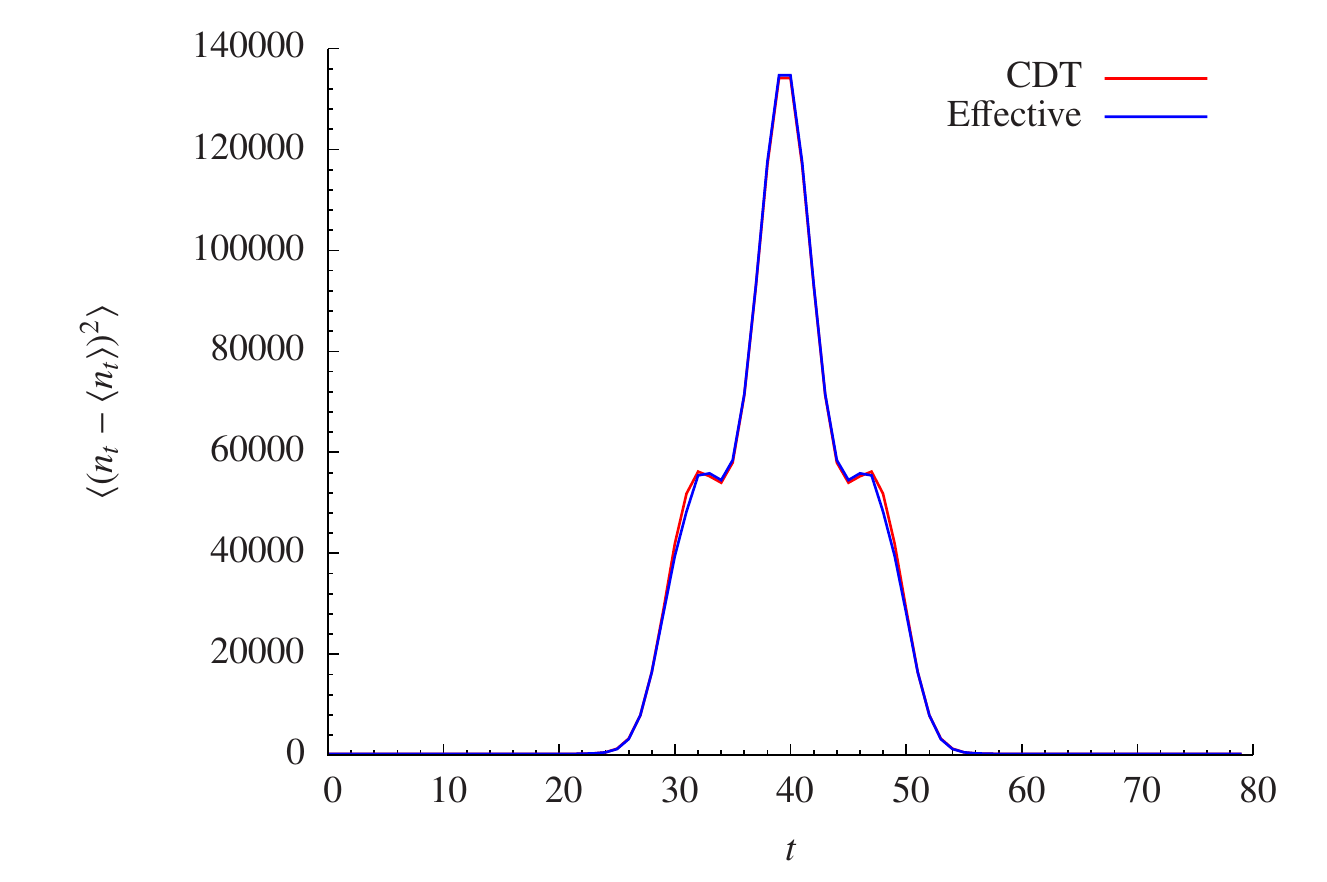}
\end{center}
\caption{Variance of spatial volumes $\langle (n_t - \langle n_t \rangle)^2 \rangle$ 
for full CDT in the de Sitter phase (red line) and the effective model (blue line).
It corresponds to the diagonal of the covariance matrix $C_{t t'}$.}
\label{fig:cov}
\end{figure}

{\it We now introduce an effective model 
which aims to reproduce results of the full CDT model in the de Sitter phase.}
In this approach configurations are given by volume profiles 
$\{ n_t \}$ rather than by triangulations $\cT$.
The model is based on the effective transfer matrix 
decomposition (\ref{Pn1nttot})
with the usual total volume fixing term used in the full CDT simulations:
\begin{equation}
 P (n_1, \dots, n_{t_{tot}}) \propto \langle n_1 | M | n_2 \rangle 
\langle n_2 | M | n_3 \rangle \cdots \langle n_{t_{tot}} | M | n_1 \rangle \ 
e^{- \epsilon \left(\sum_t n_t - \bar{N}_{41}\right)^2} \ .
\label{eq:pne}
\end{equation}
We can specify the probability distribution of configurations 
using the transfer matrix $\langle n | M | m \rangle$ 
constructed in (\ref{Mextr}).
In order to recover results of the original model we have to 
access matrix elements for large volumes.

Next, we apply standard Monte Carlo methods to generate the 
configurations, i.e. ${t_{tot}}$-component vectors 
$\{n_t, \ t = 1 \dots {t_{tot}}\}$,
according to the volume distributions (\ref{eq:pne}).
We use the same number of slices ${t_{tot}}$ and total volume $\bar{N}_{41}$ 
as in the full CDT simulations.
As before, we measure the average volume profile 
$\langle n_t \rangle$ and the covariance matrix 
$C_{t t'} \equiv \langle (n_t - \langle n_t \rangle) (n_{t'} - 
\langle n_{t'} \rangle) \rangle$.
The results obtained using this effective model are almost identical 
to the results obtained with the original, full CDT model when 
we are well into phase 'C' (the modification of Eq.\ \rf{Seffform}
needed in phase 'A' and 'B' will be discussed in Sec.\ 5 and Sec.\ 6)
Fig. \ref{fig:avt} shows the average 
volume profile $\langle n_t \rangle$ measured in the full CDT simulations 
(the red line) and in the reduced model simulations (the blue line).
The two curves overlap almost exactly.
The diagonal of the covariance matrix $C_{tt}$, 
i.e. variance of $n_t$, is shown in Fig. \ref{fig:cov}.
Again, results of the original, full CDT model (the red line) and of  
the effective model (the blue line) are in complete agreement.

\section{Extracting the  kinetic and potential terms}

The  transfer matrix measured in computer simulations can 
be used to determine the form of the effective action/Lagrangian. 
It was shown in  \cite{TM1} that in  the de Sitter phase 'C' the action 
is  approximated very well by a simple discretization of the 
continuum minisuperspace action with
a  minor small-volume correction, more precisely by the 
$L_{eff}(m,n)$ given in (\ref{Seffform}). 
The form of the discretization suggests that
 the transfer matrix can be factorized into  a potential and  a kinetic part:
\begin{equation}
\braket{n|M|m}= {\cal  N} \underbrace{ \exp  \bigg( -v[n+m] \bigg) }_{potential} \underbrace{\exp \bigg( -\frac{(n-m)^2}{k[n+m]}\bigg) }_{kinetic} \ ,
\label{SM}
\end{equation}
where  the functions: 
\begin{equation}
v[n+m]=\frac{\mu}{\Gamma} \left(\frac{n+m}{2}\right)^{1/3} - \frac{\lambda}{\Gamma}  \left( \frac{n+m}{2} \right) - \frac{\delta}{\Gamma}  \left( \frac{n + m}{2}\right)^{- \rho} 
\label{vnm}
\end{equation}
\begin{equation}
k[n+m] = \Gamma \! \cdot \! (n+m-2 n_0) \ ,
\label{knm}
\end{equation}
will be called the potential and kinetic coefficients, respectively.

The potential part can be easily analyzed by looking at the diagonal 
elements of the transfer matrix:
\begin{equation}
v[2n] = -\log{\braket{n|M|n}}+\log{\cal N} \ ,
\label{Vc}
\end{equation}
while the kinetic term requires extracting the cross-diagonal elements: 
\begin{equation}
\braket{n|M|s-n}= {\cal N}(s)\exp \left( -\frac{(2n-s)^2}{k[s]}\right) \ .
\label{Skin}
\end{equation}
By measuring the potential coefficient for different $n$ and the 
kinetic  coefficient for different $s$ one verifies  that  
eqs.\ (\ref{vnm}) and (\ref{knm}) hold in the de Sitter phase \cite{TM1}.
 
We will  apply the same factorization to analyze the 
measured transfer matrices in phases 'A' and 'B'. Further, 
we will  check how the kinetic and potential terms change when we 
move between phase 'A'  and 'C', as well as between phase 'B' and 'C'.

\section{The transfer matrix in phase 'A'}

Phase 'A' is separated from phase 'C' by a first order 
phase transition, which we meet it if we start in phase 'C' and increase 
the coupling constant $K_0$ (see Fig.\ \ref{Figfazy}).

We measured the transfer matrix in a generic point inside phase 'A' 
($K_0= 5.0, \Delta =0.4, K_4=1.22$) using the method described in 
Section 2 with $t_{tot}=2$ and a quadratic volume fixing term. 

\begin{figure}[h!]
\centering
\scalebox{0.7}{\includegraphics{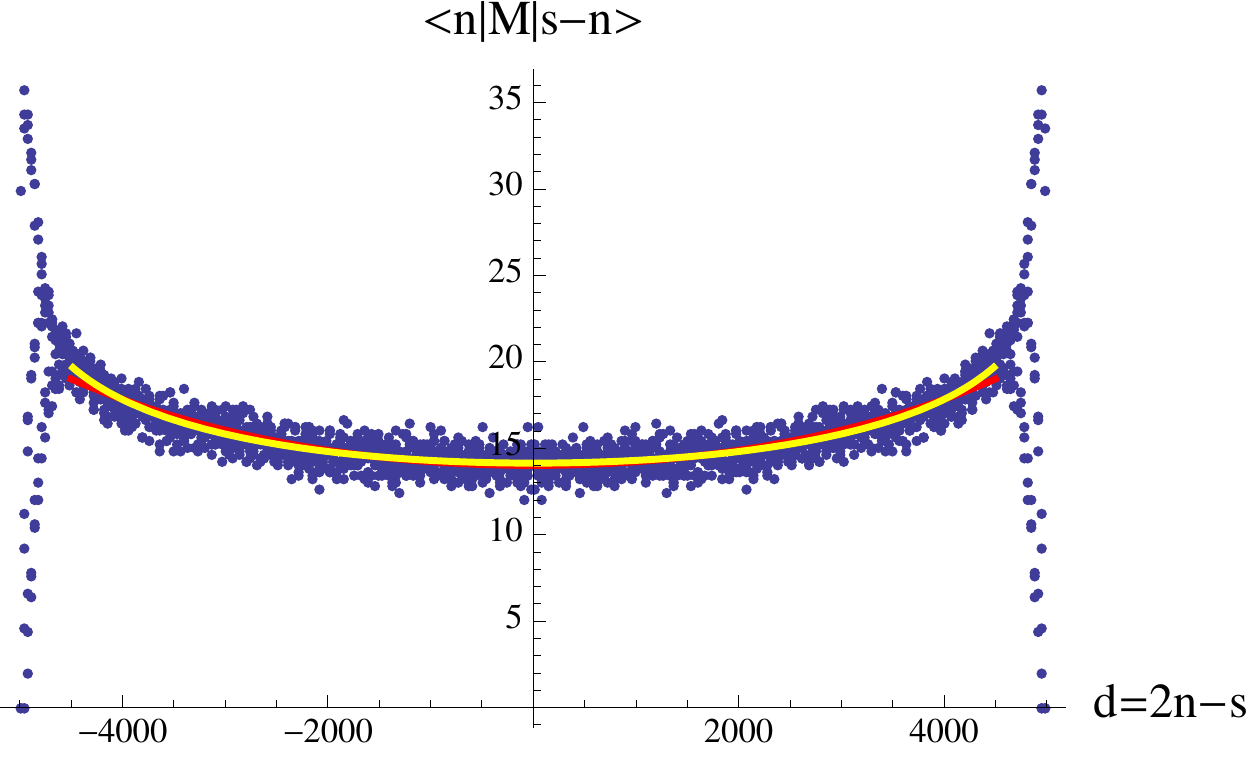}}
\caption{Sample cross-diagonal of the measured transfer matrix inside 
phase 'A' ($K_0= 5.0, \Delta =0.4, K_4=1.22$). The data are for s=n+m=5000. 
The red line corresponds to the fit of the 'artificial' ``anti-Gaussian'' 
(\ref{Akinfit}). The  yellow line is the best fit 
to the the effective Lagrangian (\ref{LeffAcross}). }
\label{Fig1}
\end{figure}

\begin{figure}[h!]
\centering
\scalebox{0.7}{\includegraphics{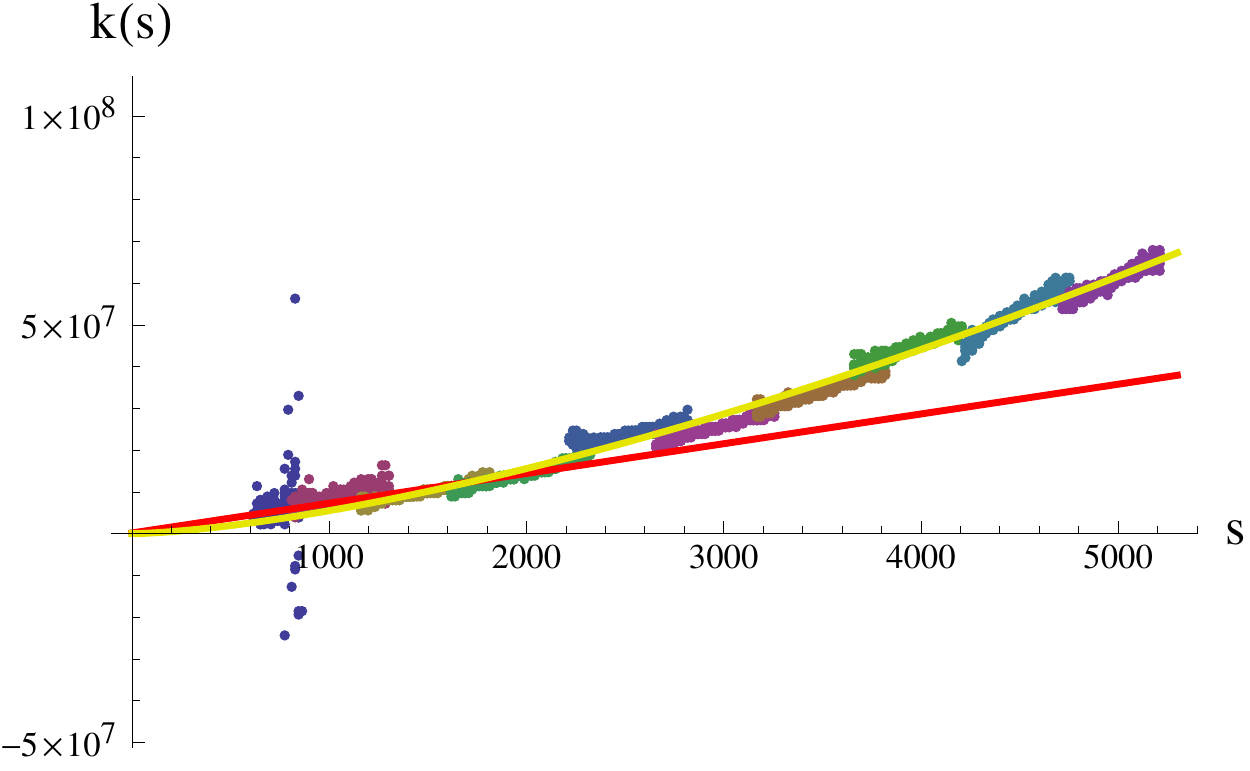}}
\caption{The kinetic coefficient $k[s]$ measured in phase 'A' 
($K_0= 5.0, \Delta =0.4, K_4=1.22$). $k[s]$ is not linear (red line) 
but can be fitted using the function defined by Eq.\  (\ref{KSa}) 
(yellow line).}
\label{Fig2}
\end{figure}

The kinetic part can be analyzed by looking at cross-diagonal elements 
of the transfer matrix: $\braket{n|M|m}=\braket{n|M|s-n}$. 
The generic  shape of the measured cross-diagonal is presented in 
Fig.\ \ref{Fig1} where $\braket{n|M|m}$ is plotted as a function 
of $(n-m)=d$. The shape looks very different from the typical 
behaviour in phase 'C' (where it  is Gaussian, cf. Fig. \ref{fig:semi}). 
Disregarding strong discretization effects for small volumes one 
could naively say that the cross-diagonals of the measured transfer matrix 
can be fitted with a very flat ``anti-Gaussian'' function. 
Indeed we tried to fit (red line in Fig. \ref{Fig1}):
 \begin{equation}
 \braket{n|M|s-n}= {\cal N}(s)\exp \left( \frac{d^{ 2}}{k[s]}\right) \ .
 \label{Akinfit}
 \end{equation}
The kinetic coefficient $k[s]$ as a function of $s$ is 
presented in Figure \ref{Fig2}.
In contrast  to the behaviour in phase 'C', $k[s]$ is no longer linear. 
It can be fitted with the following parametrization (yellow curve in  Fig. \ref{Fig2})
\begin{equation}
k[s]=k_0 s^{ 2-\alpha}.
\label{KSa}
\end{equation}
The best fit is for $\alpha=0.50\pm0.01$ and $k_0=175\pm10$. 

The ``anti-Gaussian" behaviour of the kinetic part is somewhat  strange.  
Covariance analysis of triangulations in phase 'A' shows that 
the volume distributions in different time slices are not correlated. 
Therefore the kinetic part should vanish and what we observe may just be 
an artifact of the measurement/parametrization method. 

Let us assume  that in phase 'A' the 3-volume distributions 
at different time slices are independent, but that we have a
local potential term. 
This  leads naturally to an effective Lagrangian of the  form:
\begin{equation}
L_A(n,m)= \mu \left(n^{ \alpha} + m^{ \alpha} \right) - \lambda(n+m) \ .
\label{LeffA}
\end{equation}
One can change this parametrization to: $s=n+m$, $d=n-m$ 
and assume $d/s$ small\footnote{For cross-diagonal terms it is not always 
true, as $d/s$ can be of order 1 and higher order corrections should be 
taken into account.}. Then we obtain
\begin{equation}
L_A(n,m)= \mu \left(\frac{s}{2}\right)^\alpha \left[ \left(1 + \frac{d}{s}\right)^\alpha+ \left(1 - \frac{d}{s}\right)^\alpha \right]-\lambda s =
\label{LeffAseries}
\end{equation}
$$
= -\lambda s + \mu \left(\frac{s}{2}\right)^\alpha \left[ 2 + \alpha(\alpha-1)\left(\frac{d}{s}\right)^2\right] +{\cal O}(d^4).
$$
For $\alpha<1$ we effectively get an ``anti-Gaussian'' behaviour of 
the transfer matrix cross-diagonals (\ref{Akinfit}) with
\begin{equation}
k[s] = \frac{2^\alpha}{\mu \ \alpha (	1-\alpha)}s^{ 2-\alpha},
\label{kseff}
\end{equation}
exactly in line with our measurements (\ref{KSa}). 
From the fitted values of $\alpha$ and $k_0$ 
one can calculate $\mu=0.032\pm0.002$. 

\begin{figure}[h!]
\centering
\scalebox{0.7}{\includegraphics{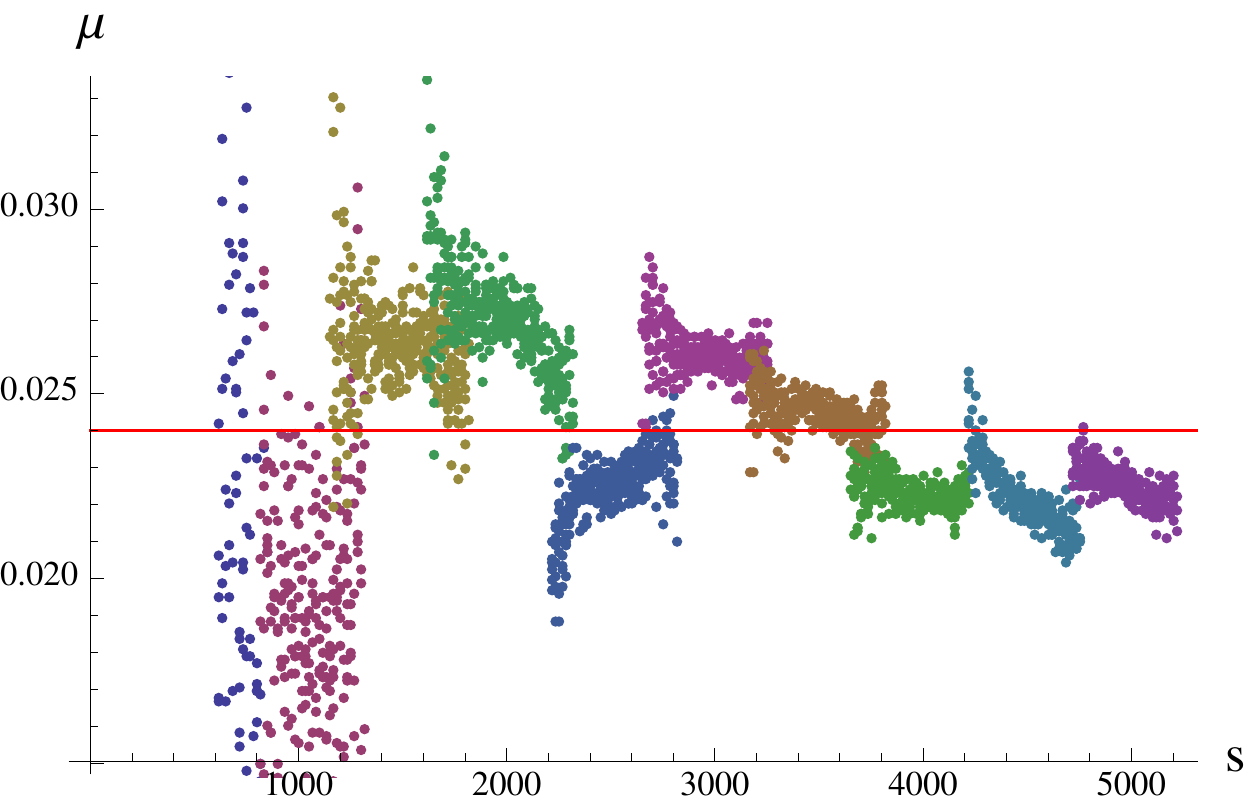}}
\caption{$\mu$ as a function of $s$ measured in phase 'A' 
($K_0= 5.0, \Delta =0.4, K_4=1.22$). The value of $\mu$ 
stabilizes around 0.024 (the red line) as  discretization effects vanish.}
\label{Fig3}
\end{figure}

As a check of parametrization (\ref{LeffA}) one may use 
the effective Lagrangian $L_A$ to fit cross-diagonal elements 
of the measured transfer matrix:
\begin{equation}
\braket{n|M|s-n}={\cal N} \exp\left[ - L_A(n,s-n)\right]={\cal N}(s) \exp\left[ - \mu n^{ \alpha} - \mu (s-n)^{ \alpha} \right] \ .
\label{LeffAcross}
\end{equation}
The best fit for $\alpha=0.5$ is presented as a yellow curve in 
Fig. \ref{Fig1} and gives $\mu=0.022\pm0.001$. 
The parameter $\mu$ fitted for different cross-diagonals 
(as a function of $s$)  is  presented in Figure \ref{Fig3}. 
The value of $\mu$  tends to a constant for large volumes (big $s$) 
as discretization effects get smaller. The  red line corresponds to $\mu=0.024$.

\begin{figure}[h!]
\centering
\scalebox{0.7}{\includegraphics{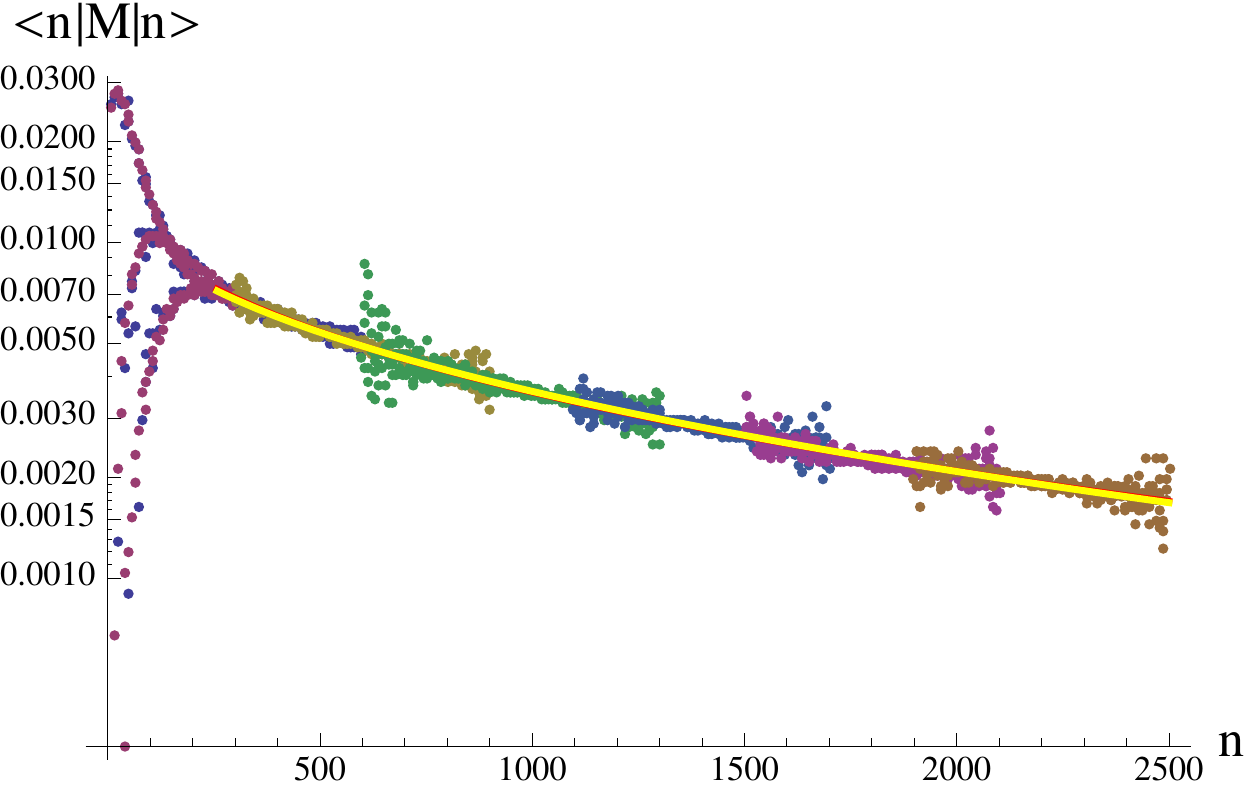}}
\caption{The diagonal elements of the transfer matrix measured 
in phase 'A' ($K_0= 5.0, \Delta =0.4, K_4=1.22$) shown 
together with the best fit to Eq.\ (\ref{MpotA}). 
The fit disregards strong discretization effects visible for small volumes.}
\label{Fig4}
\end{figure}

The  analysis of the potential part is now straightforward: 
 \begin{equation}
 \log \braket{n|M|n}=-L_A(n,n) + 
\log{\cal N}=-2 \mu n^\alpha + 2 \lambda n+ \log{\cal N} \ .
 \label{MpotA}
 \end{equation}
The diagonal elements of the measured transfer  matrix together 
with the best fit of $\mu=0.026$  (for $\alpha=0.5$)
are presented in Figure \ref{Fig4}.

As a side remark we may go back to the analysis of the de Sitter phase 'C'. 
In this phase we use the parametrization (\ref{Seffform}) 
with symmetrized potential terms. 
As a result the potential coefficient is a function of the sum of 
volumes in the adjacent time slices ($v[n+m]$). 
If instead the true potential was not symmetric 
(a function of $n$ and $m$ separately) one should expect the same 
kind of effective ``anti-Gaussian'' term to appear. However 
this effect is very small compared to the generic Gaussian 
behaviour of the kinetic part. As a result, the kinetic 
coefficients $k[n+m]$ would be slightly modified, which may 
explain the existence of the non-vanishing but very small $n_0$  
in the measured effective Lagrangian (\ref{Seffform}).

\section{The transfer matrix in phase 'B'}

Starting at the generic point in phase 'C' we reach phase 'B' by 
decreasing $\Del$ (see Fig.\ \ref{Figfazy}). 
Phase 'C' and phase 'B' are separated by a second or higher
order phase transition.

The analysis of the transfer matrix in phase 'B' is not straightforward. 
Generic triangulations in this phase are 'collapsed' i.e. the spatial
3-simplices of almost all  $\{4,1\}$ simplices
are concentrated in a single time slice. 
As a result we do not have much information about volume-volume correlations.

\begin{figure}[h!]
\centering
\scalebox{0.65}{\includegraphics{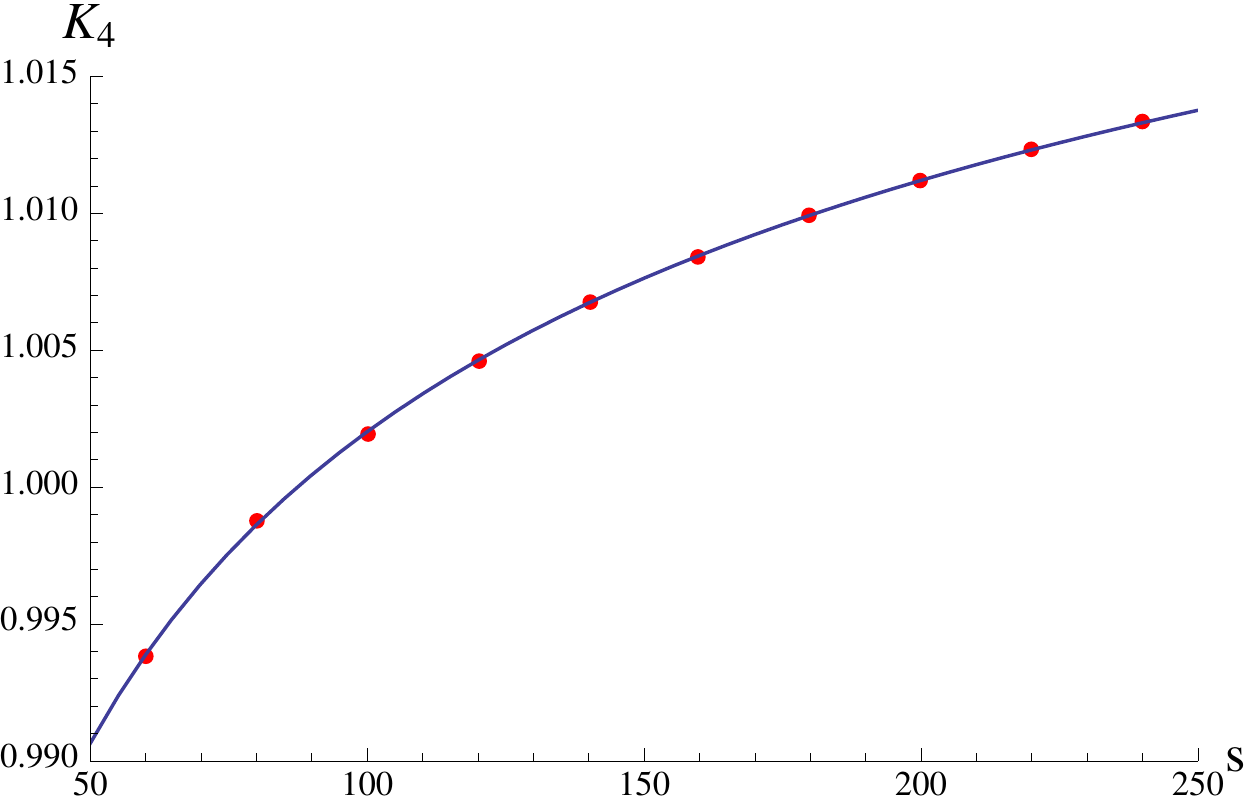}}
\caption{$K_4$ scaling with total volume (s in '000) inside phase 'B' (for $K_0=2.2 , \ \Delta=0.0$)  and the best fit of Eq. (\ref{K4vol}). }
\label{Fig4a}
\end{figure}

As an additional issue, the potential part of the effective action  
inside phase 'B' seems to suffer from a  strong non-linear 
dependence on the total volume $s$  in the small to medium volume regime. 
In our simulations we fix the $K_0$ and $\Delta$ coupling constants 
of the Regge action (\ref{Sdisc}),  while $K_4$ 
(which is conjugate to the  total four-volume of the triangulation)  
is fine-tuned to offset the spontaneously emerging entropic / potential term
coming from the exponentially large number of configurations with 
constant total volume $s$. 
If the emerging effective potential is  linear (this is the 
exponentially growing number of configurations with $s$) 
 the fine-tuned value of $K_4$ will be constant, independent of $s$. 
The non-linear corrections to the effective potential reflect  
sub-leading  corrections to the exponentially  growing number of configurations.
They might  be small power-like corrections which can effectively be 
neglected. Our simulations show that
corrections to  $K_4$ due to non-linear components  
in the effective action (\ref{Seffform}) in phase 'C'  and 
in the effective action (\ref{LeffA}) in phase 
'A'  indeed are  negligible, even in the small volume region 
(they change the fourth significant digit, which is of the same 
order as the accuracy of the $K_4$ fine-tuning). 
The situation is much different in the 'B' phase where 
the fine-tuned value of $K_4$ is strongly volume dependent 
even for relatively large volumes. It is illustrated in Fig.\ \ref{Fig4a}, 
where the value of $K_4$ is plotted as a function of total volume $s$ 
together with the fit:
\begin{equation}
K_4(s)=K_4^\infty  - \beta s^{-\gamma} \ .
\label{K4vol}
\end{equation}

The strong volume dependence of $K_4$ on $s$ implies that it is
technically impossible to measure the transfer matrix 
in phase 'B'  for the values of $K_4$ appropriate for a large volume limit. 
If we fix the $K_4$ value to the critical value corresponding to 
a large volume, effectively this value is ``too large'' and 
the system will oscillate around the minimally allowed configuration and 
only very seldom make detours to the large values of $s$ corresponding 
to the chosen value of $K_4$.
To circumvent this problem we decided to use  lower values of $K_4$ and analyze 
how a change of  $K_4$ affects the measured transfer matrix. 
As a result we can (at least qualitatively) estimate 
the properties of the transfer matrix in the continuum limit.

In this section we present the results for measurement performed at a 
generic point in phase 'B' ($K_0=2.2$ and $\Delta=0.0$). 
We start our analysis with the transfer matrix measured  
for $K_4=0.943$ using $t_{tot}=2$ with quadratic total volume fixing. 
We explicitly symmetrize the data: $\braket{n|M|m}=\braket{m|M|n}$ 
even though the measured probabilities of volume distributions 
are highly asymmetric in general. This is equivalent to regaining the  
time-reflection symmetry of the transfer matrix which is strongly broken 
by generic configurations.   

\begin{figure}[h!]
\centering
\scalebox{0.55}{\includegraphics{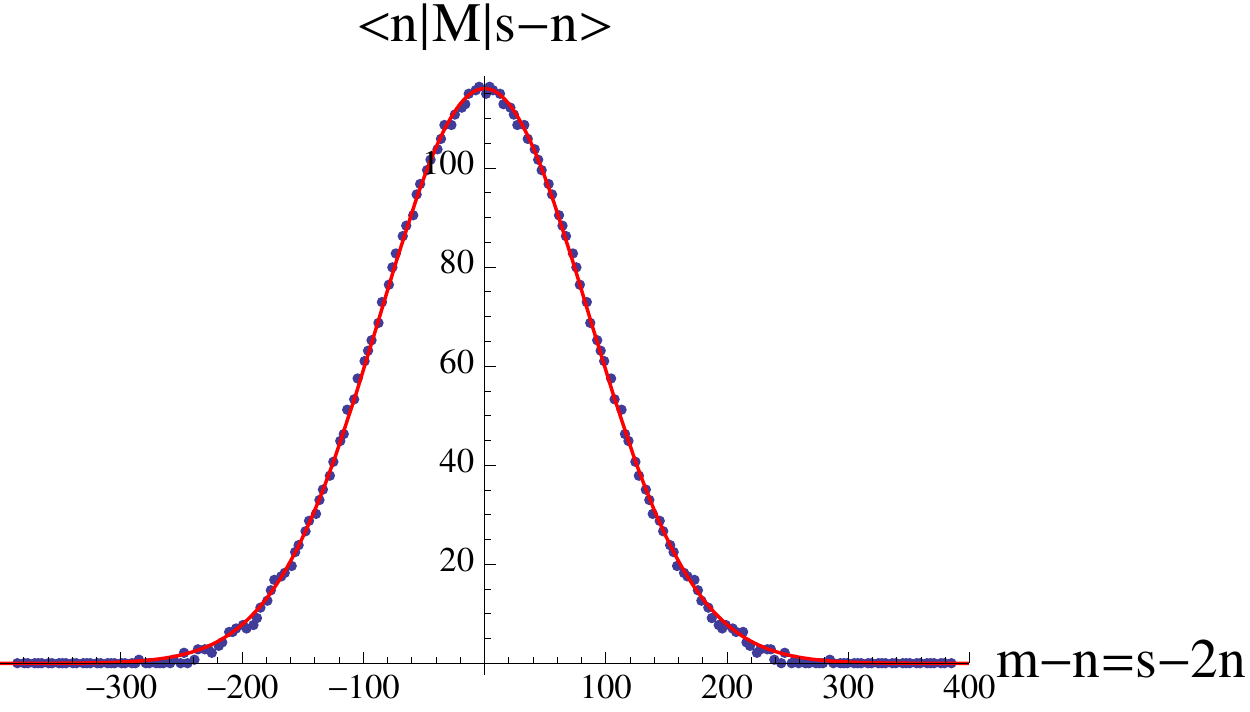}}
\scalebox{0.55}{\includegraphics{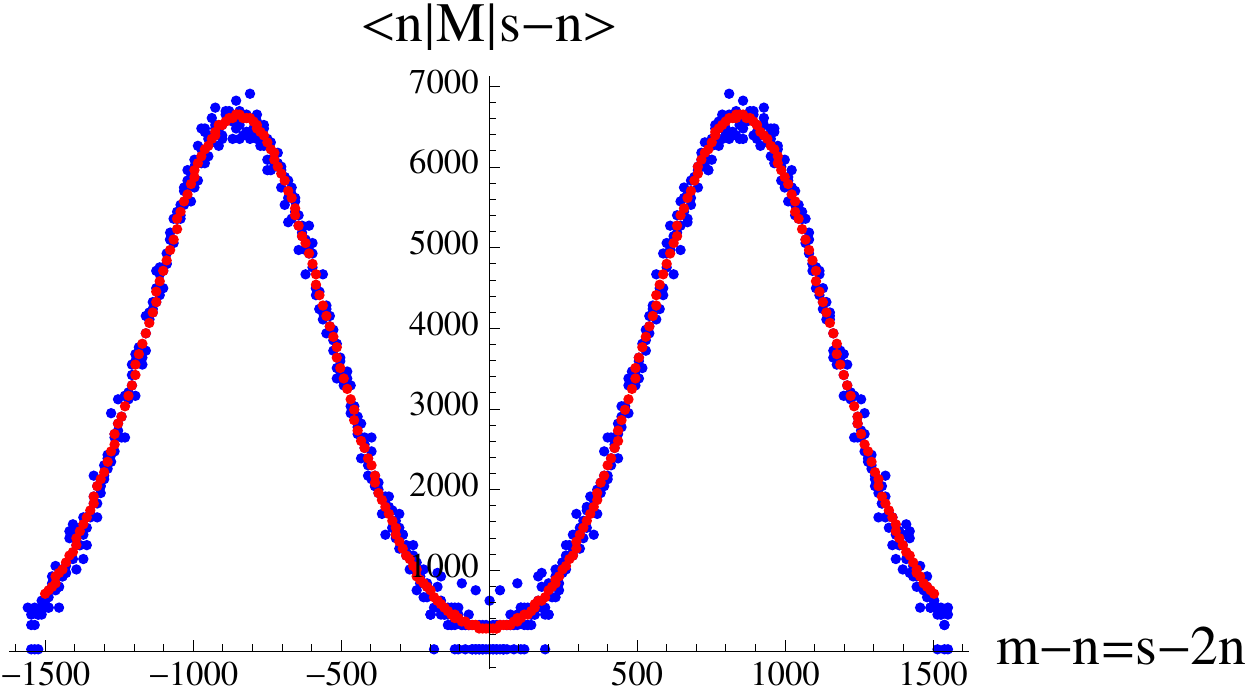}}
\caption{Cross diagonals of the transfer matrix measured in phase 'B' (for $K_0=2.2$, $\Delta=0.0$ and $K_4=0.943$). The left chart shows the data  below bifurcation point ($s=n+m<s^b$). The right chart presents cross-diagonals  above bifurcation point ($s>s^b$). The best fits of Eq. (\ref{SkinB}) are presented as red lines.}
\label{Fig5}
\end{figure}

\begin{figure}[h!]
\centering
\scalebox{0.7}{\includegraphics{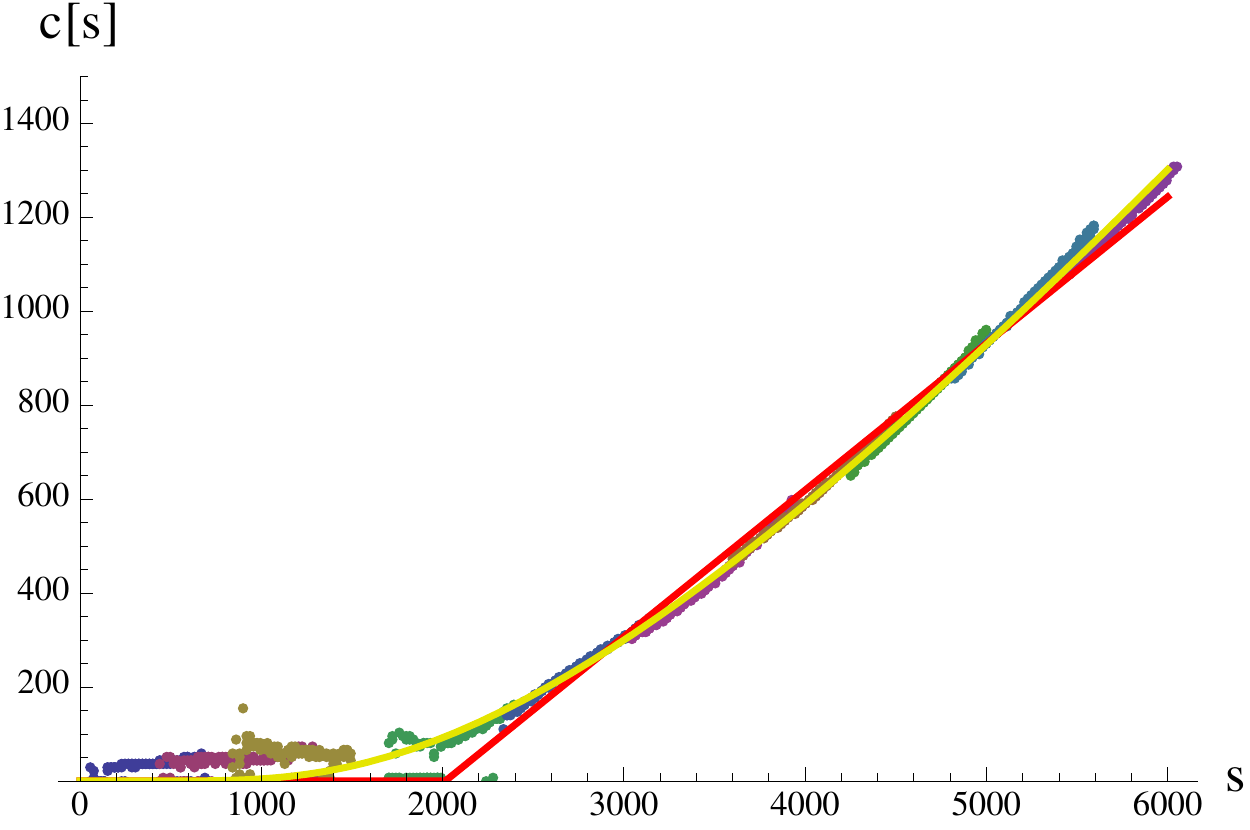}}
\caption{The bifurcation shift c[s] measured in phase 'B' 
(for $K_0=2.2$, $\Delta=0.0$ and $K_4=0.943$) 
together with the best fits using Eq. (\ref{cs}) 
(red line) and (\ref{csexp}) (yellow line).}
\label{Fig6}
\end{figure}

The typical behaviour of the cross-diagonal (kinetic) part of the 
measured transfer matrix $\braket{n|M|m}=\braket{n|M|s-n}$ 
strongly depends on $s=n+m$. For $s<s^b$ it looks the same 
as in phase 'C' and can be well fitted with a single Gaussian 
(\ref{Skin}) - see Fig. \ref{Fig5} (left). For $s>s^b$ 
the cross-diagonals split into the sum of two ``shifted" Gaussians - 
see Fig. \ref{Fig5} (right). The value of the shift depends on $s=n+m$ 
(Figure \ref{Fig6}). All together the kinetic part can be parametrized by:
\begin{equation}
\braket{n|M|m}=\braket{n|M|s-n}=
\label{SkinB}
\end{equation}
$$
 ={\cal N}(s)\left[ \exp \left( -\frac{\Big((m-n) - c[s]\Big)^2}{k[s]}\right) + \exp \left( -\frac{\Big((m-n) + c[s]\Big)^2}{k[s]}\right)\right] \ ,
$$
where: $c[s]$ is (close to) zero for $s<s^b$ and (almost) linear for $s>s^b$:
\begin{equation}
c[s] \approx \max[0,c_0(s-s^b)] \ .
\label{cs}
\end{equation}
This type of parametrization fits the measured data quite well 
(the red line in Fig. \ref{Fig6}) and is convenient for our further analysis. 
We will call $s^b$ the {\it bifurcation point}. 
For our generic data ($K_0=2.2$, $\Delta=0$ and $K_4=0.943$) 
the best fits yield: $s^b=2020$ and $c_0=0.31$.

Another phenomenological parametrization, which fits the data around bifurcation point even better (yellow curve in Fig. \ref{Fig6}) is: 
\begin{equation}
c[s] = c_0  \;s\;\exp(-s^b/s) \ .
\label{csexp}
\end{equation}
It is consistent with (\ref{cs}) for small and large $s$ (compared to $s^b$). 
We will return to this parametrization when 
analyzing the phase transitions in the next sections.

\begin{figure}[h!]
\centering
\scalebox{0.7}{\includegraphics{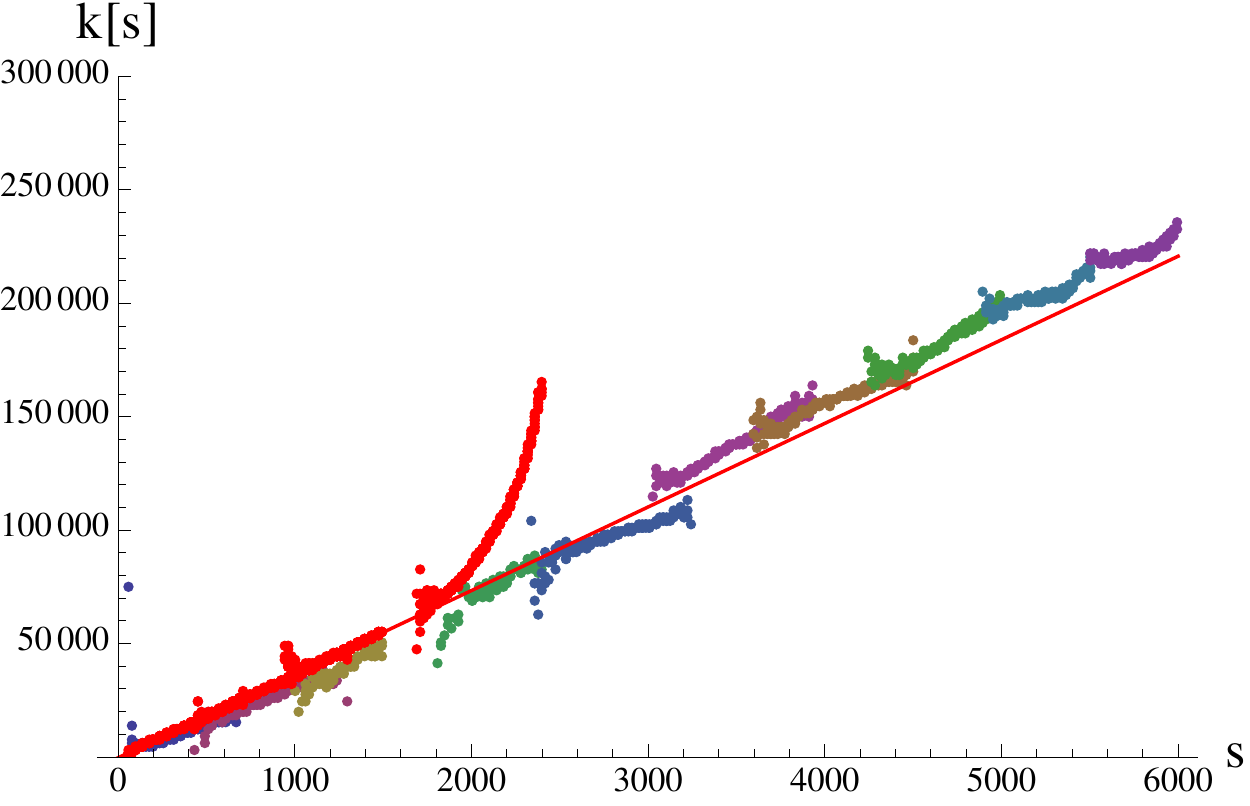}}
\caption{The kinetic coefficient k[s] 
measured in phase 'B' (for $K_0=2.2$, $\Delta=0.0$ and $K_4=0.943$).  
The red points correspond to the fit of a single Gaussian 
(which is not valid  after crossing bifurcation point (for $s=n+m>s^b$)). 
Different colours  correspond to the fit of two  Gaussians (\ref{SkinB}) 
which is reliable also above the bifurcation point (for $s>s^b$). 
The bifurcation point $s^b$ can be identified as the point 
at which the fit of the single Gaussian starts to diverge from  
linear behaviour. It is consistent with the measured value  of $s^b$ 
as presented in Fig.\ \ref{Fig6}.}
\label{Fig7}
\end{figure}

The last function that should be fitted is $k[s]$ which is very 
well approximated by a linear function, independently   
of whether we are below or above the bifurcation point (Fig. \ref{Fig7}). 
The behaviour is consequently the same as in phase 'C', i.e.\ (\ref{knm}):
\begin{equation}
k[s] = \Gamma(s - 2 n_0) \ .
\label{ksB}
\end{equation}
The best fit yields: $\Gamma= 36.8$, $n_0=5.4$ 
which is of the same order as the values measured in the de Sitter phase 'C'.

\begin{figure}[h!]
\centering
\scalebox{0.55}{\includegraphics{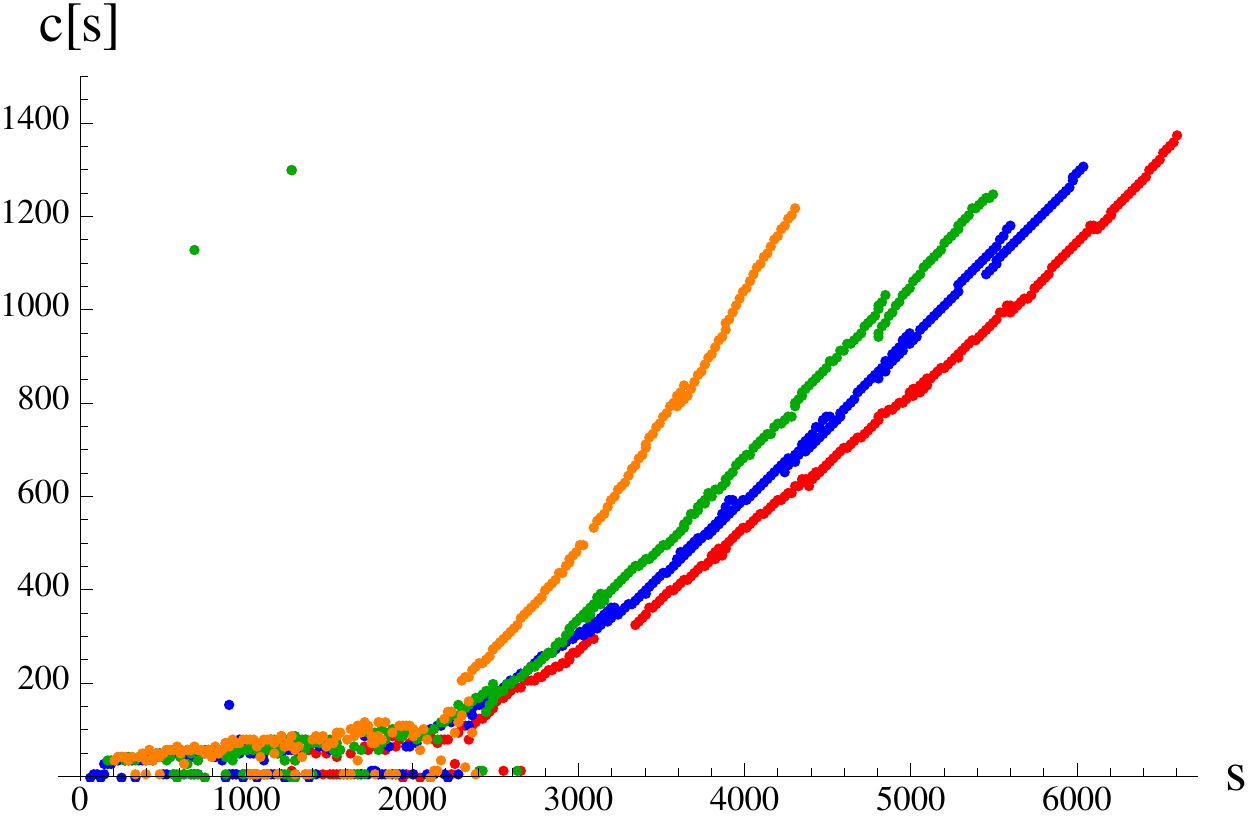}}
\scalebox{0.55}{\includegraphics{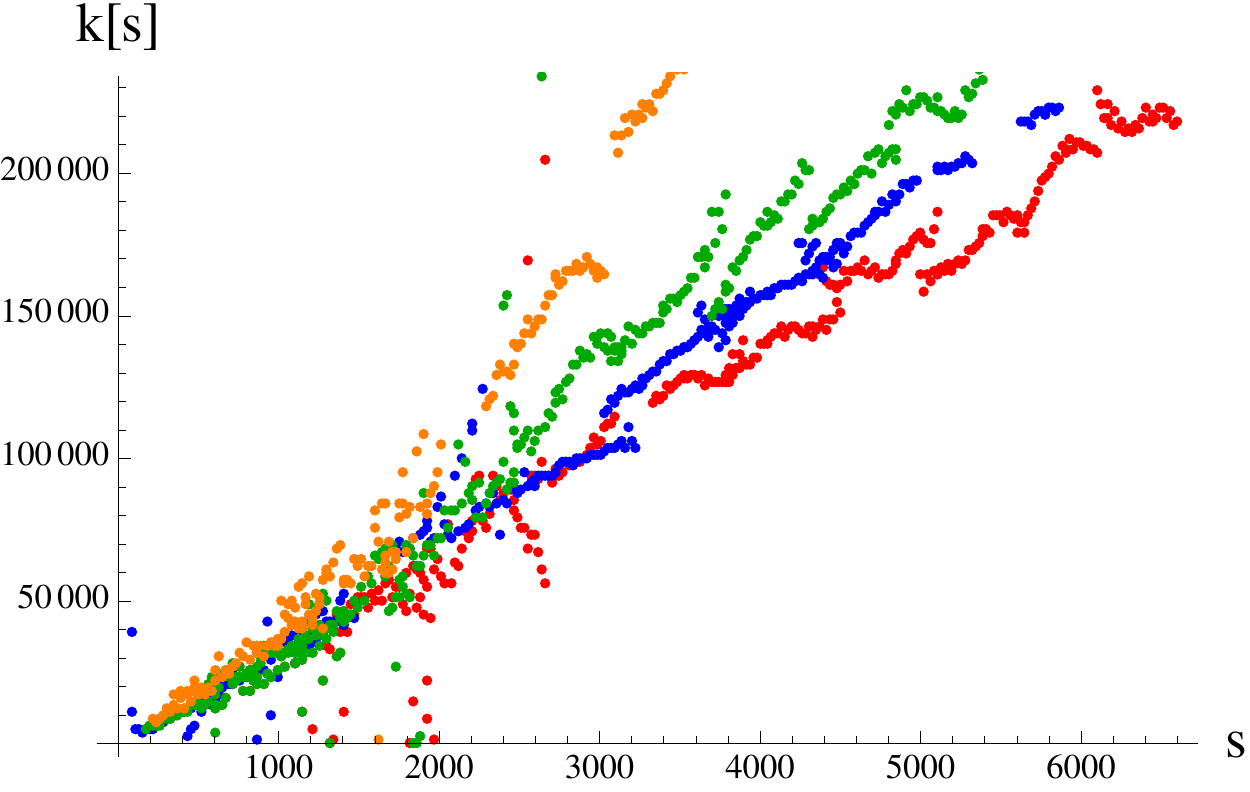}}
\caption{The bifurcation shift $c[s]$ (left) and the 
kinetic coefficient $k[s]$ (right) measured  in phase 'B' 
($K_0=2.2$, $\Delta=0.0$) for different values of  
$K_4=$ 0.933 (red), 0.943 (blue), 0.953 (green), 0.973 (orange). 
The bifurcation 
point $s^b$ is stable whereas the slope of the shift $c_0$ 
and the value of $\Gamma$ rise  as $K_4$ is increased.}
\label{Fig8}
\end{figure}

As we are interested in properties of the transfer matrix in 
the large volume limit (where critical values of $K_4$ are much higher) 
it is important to check how  the results depend on $K_4$. 
The plots of $c[s]$ and $k[s]$ for different $K_4$ are presented in 
Fig.\ \ref{Fig8}. In general, the functional form of Eq's (\ref{SkinB})-(\ref{ksB}) 
is adequate for different values of $K_4$.  With regards to the parameters
entering in Eq's (\ref{SkinB})-(\ref{ksB}), the change of $K_4$ 
does not influence the position of the bifurcation point $s^b$,
while the bifurcation slope $c_0$ and the effective Newton's constant 
$\Gamma$ rise as $K_4$ is increased. 

Let us use these results to explain (at least qualitatively) 
the behaviour of the system in phase 'B'. In our 'full CDT' simulations 
(with $t_{tot}=80$)  we analyze systems with large total volume 
($\geq 40$k simplices) for which $K_4$ is big. 
As a result  $c_0$, and consequently $c[s]$, are large 
in the interesting region ($s>s^b$). Naively speaking, 
configurations with very large difference of spatial volume 
in the adjacent time slices should be favoured  
(most probable $(m-n)$ is large) and a kind of 'anti-ferromagnetic' 
behaviour with a ...-'large'-'small'-'large'-'small'-.... volume 
distribution observed. This is exactly what we see  in 
CDT systems with small time periods $t_{tot}=2, 4, 6$ 
used in the transfer matrix measurements, but for $t_{tot}=80$ 
the observed behaviour is very different and the volume distribution 
is 'collapsed' to just one time slice. In order to explain 
this phenomena we must take into account the entropic factor  
(the potential part in the transfer matrix 'language'). Due to 
strong dependence of $K_4$ on the total volume, the exact measurement 
of the potential in the large volume range is beyond our reach at the moment. 
Instead let us present a theoretical model in which the potential 
is exactly the same as in phase 'C', i.e.\ given by Eq.\ (\ref{vnm}). 
For simplicity we will consider only the leading behaviour 
by setting  $\lambda, \delta, n_0 = 0$. Consequently:
$$
\braket{n|M|m} =
\exp \left( - \frac{\mu}{\Gamma} \left(\frac{n+m}{2}\right)^{1/3} \right)
\left[ 
\exp \left( -\frac{\Big((m-n) -\big[ c_0(n+m-s^b) \big]_+\Big)^2}{\Gamma(n+m)}\right) + \right.
$$
\begin{equation}
\left.
+\exp \left( -\frac{\Big((m-n) +\big[ c_0(n+m-s^b) \big]_+\Big)^2}{\Gamma(n+m)}\right) 
\right] \ ,
\label{bifmodel}
\end{equation}
where: $[.]_+ = \max(. , 0)$.

\begin{figure}[h!]
\centering
\scalebox{0.9}{\includegraphics{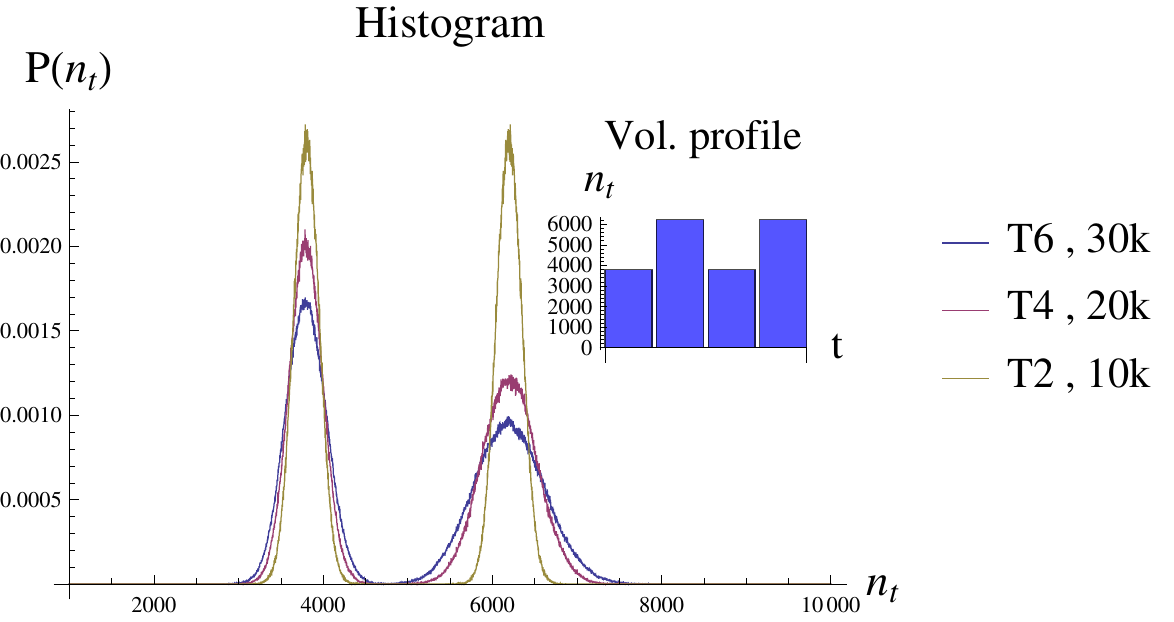}}
\caption{The histogram of the spatial volume distributions and the volume profile
measured in the effective Monte Carlo model (\ref{bifmodel}) 
for $c_0=0.3$ and $t_{tot}=2,4,6$. The two Gaussian peaks 
correspond to odd and even time slices, respectively. As a result 
the average volume profile is 'anti-ferromagnetic' with quantum 
fluctuations around: ...-3.8k-6.2k-3.8k-6.2k-... .}
\label{Figeffmodel}
\end{figure}

\begin{figure}[h!]
\centering
\scalebox{0.9}{\includegraphics{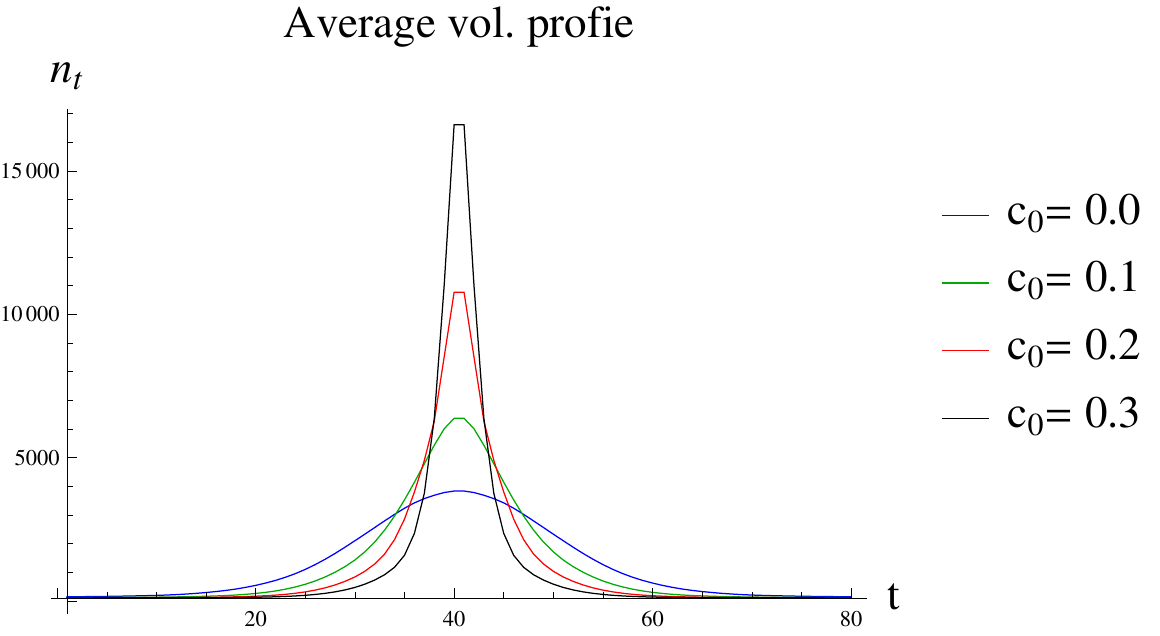}}
\caption{The average spatial volume 
measured in the effective Monte Carlo model (\ref{bifmodel}) 
for $t_{tot}=80$, 
$\bar N_{41}=100$k and different values of $c_0$. 
The shape of the volume profile is consistent 
with the 'collapsed' blob structure for $c_0 >0$.}
\label{Figeffmodel2}
\end{figure}

Now we can perform the same kind of 'effective' Monte Carlo 
simulations as explained in section 3 in which the theoretical 
transfer matrix (\ref{bifmodel})  will be used to generate 
volume distributions $\{ n_t, t=1...t_{tot}\}$ with the probability 
given by Eq.\ (\ref{eq:pne}). We set the parameters of our model 
to  the  values measured in the real 'full CDT' simulations: 
$\Gamma=37, \mu=15, s^b=2000$ and  $c_0 = 0.1 - 0.3 $.  
The resulting volume distribution for small and large $t_{tot}$ 
is presented in Fig. \ref{Figeffmodel} and Fig. \ref{Figeffmodel2}, respectively. As a reference case we 
also plot the volume distribution for $c_0=0$, for which we 
recover the generic behaviour found in the de Sitter phase 'C'.
 
For small $t_{tot}$  
the expected 'anti-ferromagnetic' structure is observed, 
while for large $t_{tot}$ a single 'collapsed' blob forms. 
The strength of the 'collapsed' behaviour depends on $c_0$. 
This simple model explains very well (at least at a qualitative level) 
the volume distribution inside  phase  'B'. In reality  
we should  take into account that the 
value of $c_0$ appropriate for large $K_4$ used in 'full CDT' simulations 
is probably much bigger (leading to a much more narrow distribution 
for large $t_{tot}$). In addition  the  actual  entropic/potential part
present in the full CDT model may corroborate the idea of 
 a 'narrowing' of the volume distribution compared to the one 
we observe in the toy model defined by \rf{bifmodel}. 
 
\section{Phase transitions}

\begin{figure}[h!]
\centering
\scalebox{0.8}{\includegraphics{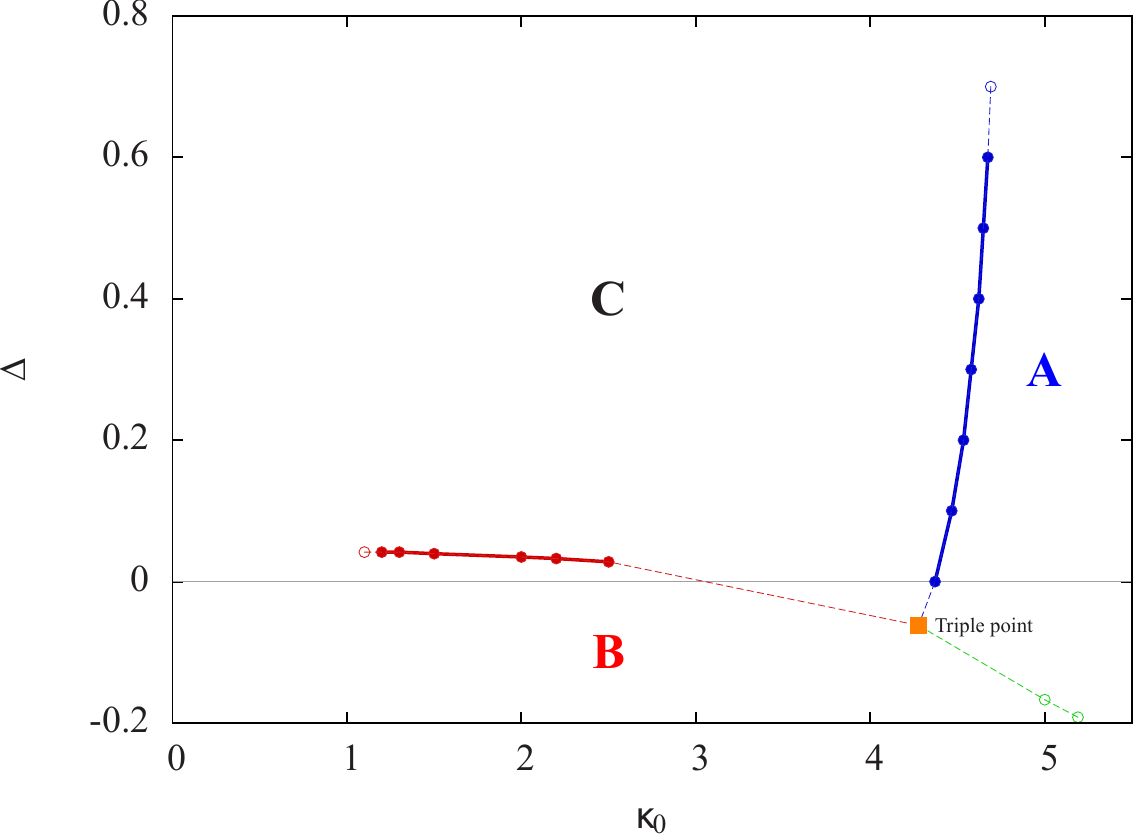}}
\caption{The CDT phase diagram measured by 
'traditional' methods based on the order parameter analysis.}
\label{Figfazy}
\end{figure}

When one applies  conventional methods to analyze the phase
transitions observed in four-dimensional CDT one obtains strong 
evidence that the 'A'-'C' transition is a first order transition 
while the 'B'-'C' transition is a second (or higher) order transition
\cite{phases1}.    
These results are based on the analysis of order parameters 
defined as variables conjugate to bare coupling constants in the 
Regge action (\ref{Sdisc}). By looking at the susceptibility of 
the order parameters one can identify the position of the 
phase transition lines in the phase diagram (Fig. \ref{Figfazy}).  
At the same time the critical exponents, scaling properties and 
large volume behaviour enable one to analyze the order of the phase transition.
The order parameters in question reflect  some global 
characteristics  of the CDT triangulations 
(e.g. the ratio $N_0/N^{\{4,1\}}$). A change in such order parameters 
does not necessarily give much insight into the 'microscopic' nature of the 
phase transitions, which is an obvious  drawback of this approach.
In particular, if one wants to find algorithms which can beat the 
critical slowing down observed near the transitions.
We will try to use  the transfer matrix to obtain additional 
information about the phase transitions.

\begin{figure}[h!]
\centering
\scalebox{0.55}{\includegraphics{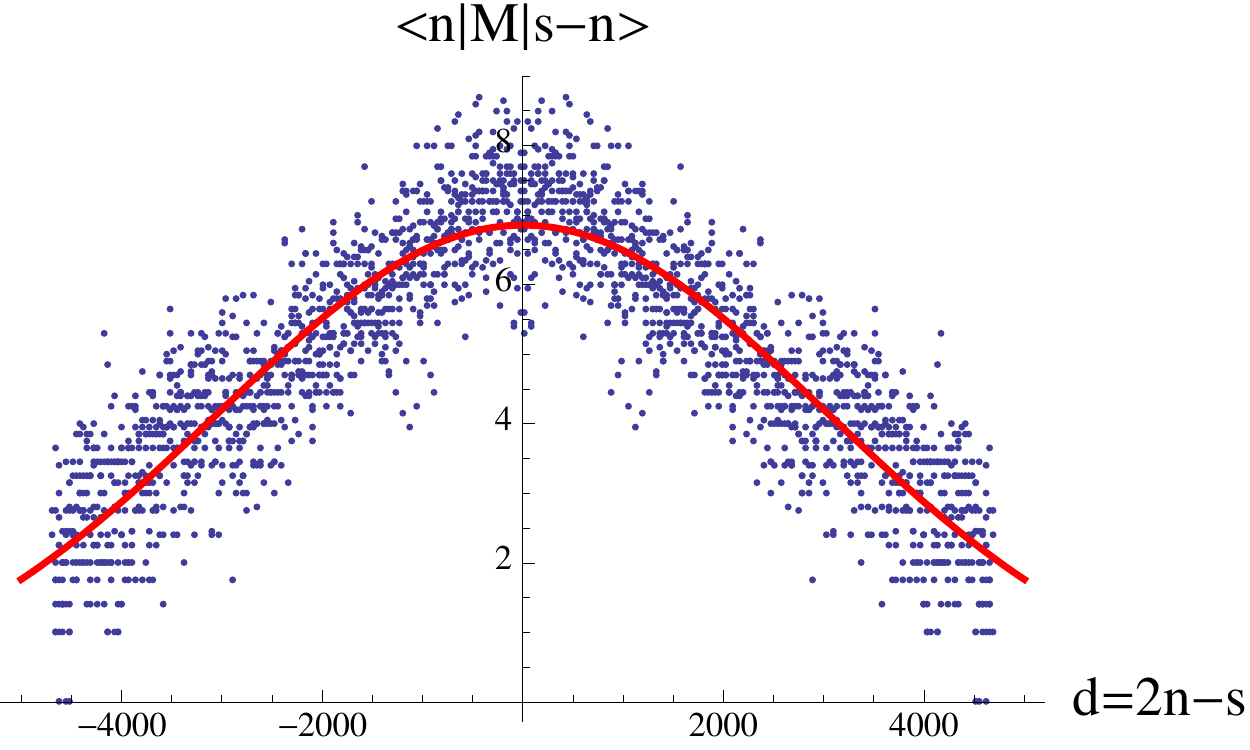}}
\scalebox{0.55}{\includegraphics{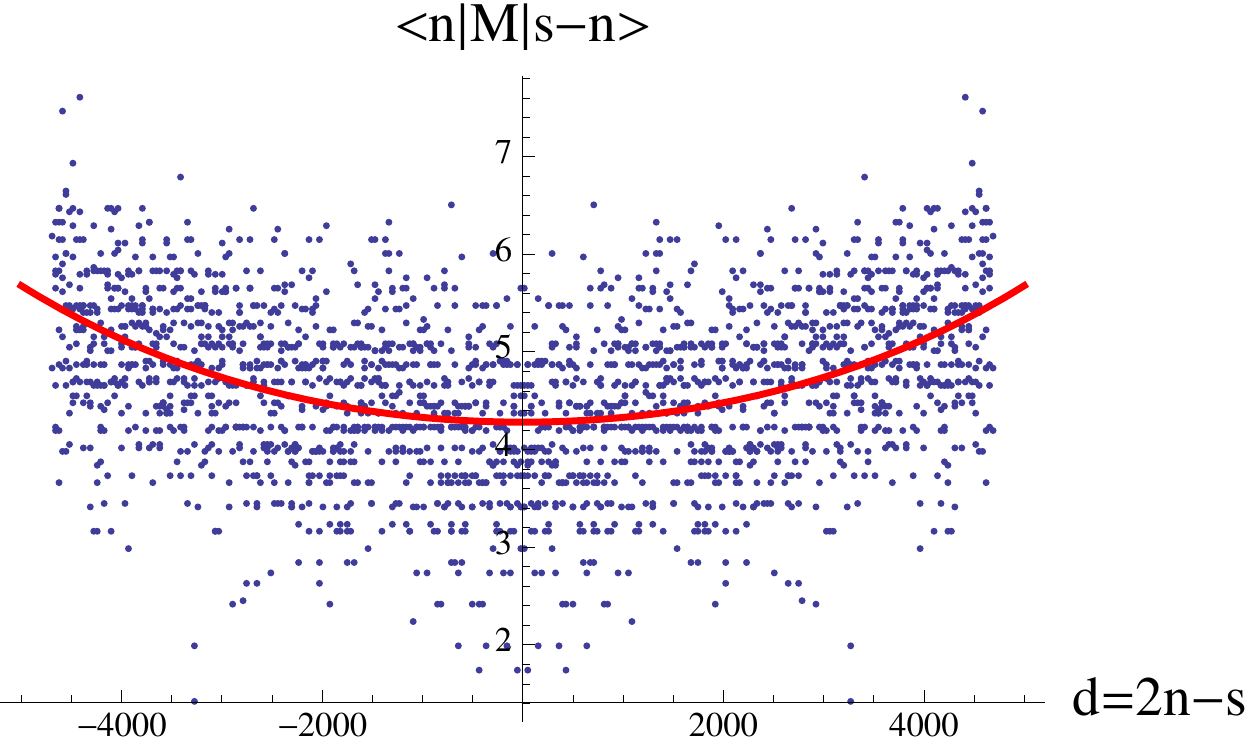}}
\caption{The cross-diagonal elements of the transfer matrix 
for $s=n+m=5000$ measured  
for $\Delta=0.6$. The left chart presents data for  
$K_0=4.7$ (phase 'C') while the right chart presents the data 
for $K_0=4.8$ (phase 'A'). The change of the behaviour 
is clearly visible, which enables us to identify the phase transition point. }
\label{Fig10}
\end{figure}

The 'A'-'C' phase transition is easily visible 
in the kinetic part of the transfer matrix. When we  approach the  'A'-'C'  phase 
transition line from phase 'C' 
(by increasing $K_0$ and keeping $\Delta$ fixed) 
the kinetic part of the minisuperspace effective action 
(\ref{Seffform}) vanishes  smoothly. 
Near an 'A'-'C'  phase transition point  the cross-diagonals  
of the measured transfer matrix are almost constant. 
Just after the phase transition  we can observe  
the formation of the 'artificial' anti-Gaussian term discussed  
in detail in section 4. For  $\Delta=0.6$ the phase transition 
point can be identified  at $K_0 = 4.75\pm 0.05$ (see Fig. \ref{Fig10}) 
which is fully consistent with the location found using 
the 'traditional'  approach used in \cite{phases1}.

\begin{figure}[h!]
\centering
\scalebox{0.7}{\includegraphics{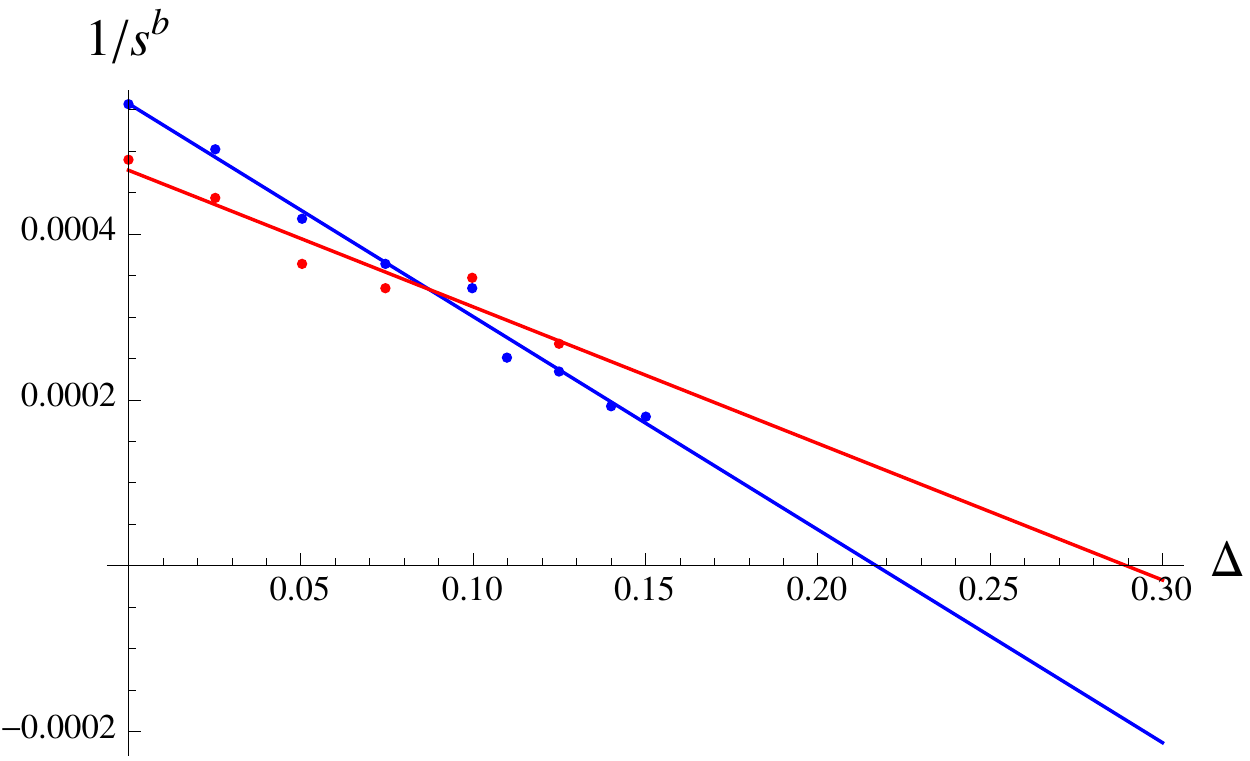}}
\caption{The (inverse of) bifurcation point $s^b$ as a function of $\Delta$ (for fixed $K_0=2.2$). The colours correspond to two different 
ways of extracting $s^b$: by direct measurements (red points) and  
indirectly, as the values of $s$  where a single Gaussian does not 
fit the distributions (blue points) - see footnote $^3$. }
\label{Fig11}
\end{figure}

The 'B'-'C' phase transition is not as easily visible. 
In the previous section we parametrized the kinetic part of the 
transfer matrix in phase 'B' by a sum of two Gaussians 
(\ref{SkinB})-(\ref{ksB}). An obvious parameter to look at 
is the bifurcation point $s^b$. For small volumes 
$s=n+m<s^b$ the kinetic part is the same as in phase 'C'. 
The difference, responsible for the 'collapsed' behaviour in phase 'B', 
is observed for large volumes $s>s^b$. Thus, it is a natural conjecture 
that the  'B'-'C' phase transition is related to the appearance 
of a bifurcation point $s^b$. However, as we will see that is not 
the case. If we start in phase 'B', keep $K_0$ fixed 
and increase $\Delta$ in order to cross the  'B'-'C' phase transition line
the value of $s^b$ also increases.
Thus, it is natural to treat the condition $s^b \to \infty$  as a sign of 
a phase transition. In Figure \ref{Fig11} we present the plot of $1/s^b$ 
as a function of $\Delta$ for $K_0=2.2$. 
Different colours correspond to two methods of measuring $s^b$. 
The relation seems to be linear, implying the  transition occurs 
for $\Delta = 0.2-0.3$. This value of $\Delta$ is much higher 
than critical value measured in the 'traditional' approach 
($\Delta \approx 0.05 $). By using $s^b$ as an indicator of 
a phase transition we are seemingly observing 
something different from the formerly observed 'B'-'C' transition.
We will discuss this  in the next section.

\section{A new 'bifurcation' phase?}

 \footnotetext[3]{Red points correspond to $s^b$ determined  by fitting Eq. (\ref{cs}) 
to the measured data. This method requires performing 
transfer matrix measurements also in the region of  volumes much higher 
then $s^b$ which is difficult as we approach the phase transition point.  
Blue points correspond to the indirect determination of $s^b$,
identified as the point at which a single Gaussian does no longer fit the  
measured cross-diagonals (recall Fig. \ref{Fig7} for details). 
The larger the $\Delta$ the more difficult it is to observe 
the shift away from a single-Gaussian distribution. 
Therefore the values of $1/s^b$ for large $\Delta$ 
are probably underestimated when using this second method.}

\begin{figure}[h!]
\centering
\scalebox{0.7}{\includegraphics{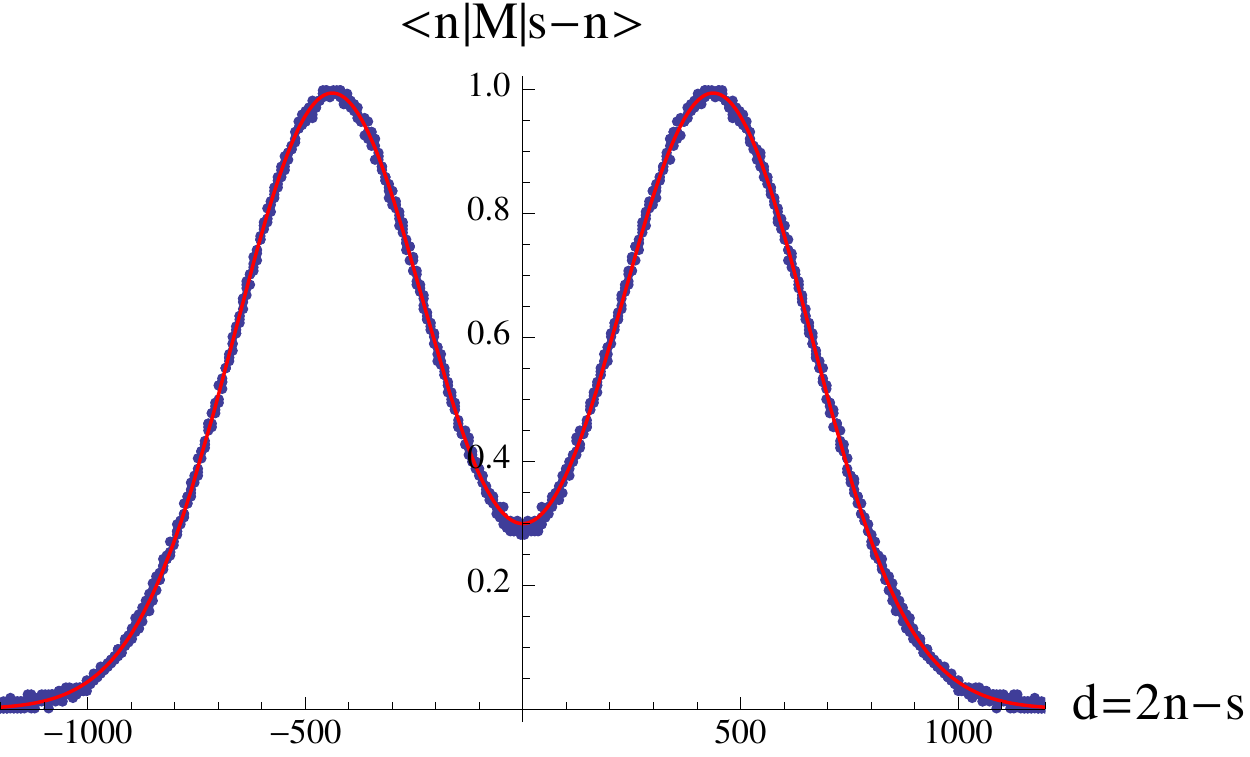}}
\caption{The cross diagonal elements of the transfer matrix measured 
for $K_0=2.2$ and $\Delta=0.125$ which according to the 'traditional' 
approach lies well inside the de Sitter phase 'C'. 
The data are measured for $s=n+m=15000$. The double-Gaussian 
bifurcation structure characteristic for phase 'B' configurations 
is still clearly visible.}
\label{Fig12}
\end{figure}

In the previous section we provided evidence that a new kind 
of transition occurs when we start out in phase 'B' and increase $\Del$.
It is related to the disappearance of the bifurcation point $s^b$ observed
in phase 'B' in the kinetic term. However, this disappearance of $s^b$  
is observed for the values of $\Delta$ much larger than the 
$\Del$-value where the 'conventional' 'B'-'C' phase transition is located.
It is thus located in the region of coupling constant space 
we  conventionally have denoted phase 'C'.
The position of the new transition point is based on the 
interpolation of the  bifurcation point $s^b$ to infinity 
as a function of $\Delta$. One may argue that 
this relation may change in the vicinity of the  transition, 
lowering the 'critical' $\Delta$ value. However, it is possible  
 to observe the 'bifurcation' structure also for $0.1<\Delta<0.3$ 
(i.e.\ in the region of the 'C' phase bordering the conventional 
'B'-'C' phase transition) if one uses a  total 
volume $\bar{N}_{41}$  large enough 
(see Figure \ref{Fig12}) and performs the simulations with small $t_{tot}$. 
For the same values of $\Del$ the average volume profile for large $t_{tot}$ 
has the typical blob-shape characteristic for the de Sitter phase 'C'.  
In fact, if one looks at the transfer matrix data nothing 
special happens while crossing the conventional 'B'-'C' phase transition line 
($\Delta \approx 0.05$ for $K_0=2.2$). This is in obvious 
contradiction with the 'traditional' phase diagram presented in 
Fig.\ \ref{Figfazy}.

To explain this phenomenon we refer back to Fig.\ \ref{Figeffmodel} 
(right) where the average volume profiles for the 
effective Monte Carlo model with the 
'bifurcated' transfer matrix (\ref{bifmodel}) 
were shown. For small bifurcation slopes $c_0$ the volume profiles 
are practically identical with those observed in the generic de Sitter phase. 
This is true even for systems with large total volumes ($\bar{N}_{41} = 100$k). 
For medium bifurcation slopes  the volume profile contracts 
in the time direction, but the general shape does not change much. 
Only for  large $c_0$ does one observe something which 
resembles a 'collapse' of the blob in the time-direction. 

\begin{figure}[h!]
\centering
\scalebox{0.9}{\includegraphics{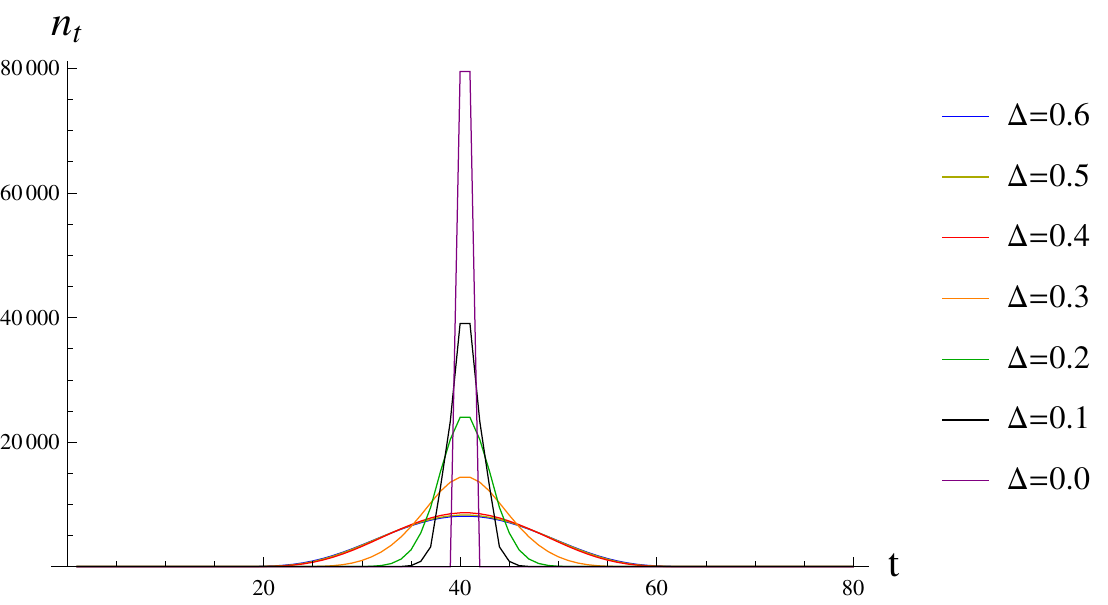}}
\caption{The average spatial volume profiles measured in 'full CDT' 
for $K_0=2.2$, $t_{tot}=80$ and $\bar{N}_{41} = 160$k.}
\label{FigavCDT}
\end{figure}

If one plots the average volume profiles measured in 'full CDT' for 
$K_0=2.2$ and different values of $\Delta = 0.0 - 0.6$ 
(Fig.\ \ref{FigavCDT}) the result looks qualitatively the same. 
For $\Delta \geq 0.4$ the shape does not change much. An increasing contraction 
in the time-diction takes place for $0.1 \leq \Delta \leq 0.3$. Finally,
crossing the 'conventional' 'B'-'C' phase transition at $\Del=0.05$ we
observe the  'collapse'  of the  blob in the time direction, characteristic 
of a  generic phase 'B' configuration.

\begin{figure}[h!]
\centering
\scalebox{0.85}{\includegraphics{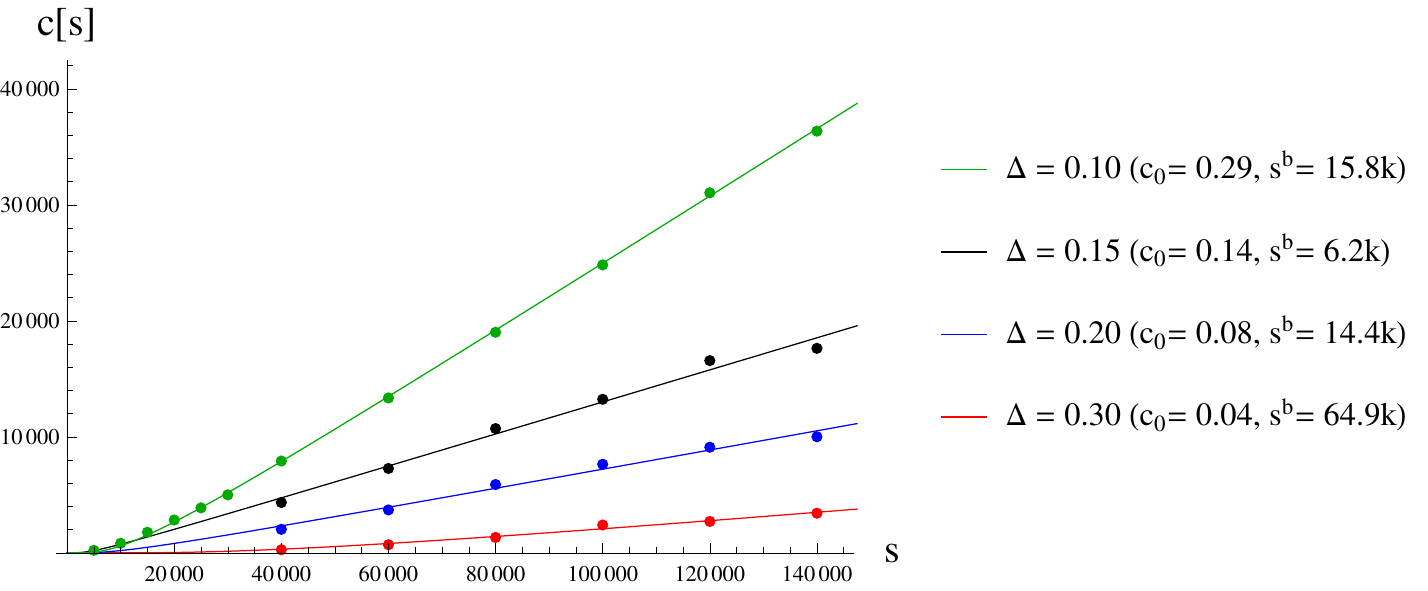}}
\caption{The bifurcation shifts (dots) measured for $K_0=2.2$ and for
different values of $\Delta$,  and the best fits of 
$c[s]=c_0 \exp(-s^b / s)\, s$ to these data (lines).   }
\label{FigC0Sb}
\end{figure}

To justify this picture we  measured the bifurcation structure 
for  $0.1 \leq \Delta \leq 0.3$. It required using large 
total volumes which substantially increased the computer simulation time.
To get better statistics we focused on selected cross-diagonals of the 
measured transfer matrix by choosing the  global volume fixing term 
(\ref{VFpot1}) very peaked at $n_{vol} = 20$k, $40$k, $60$k, $80$k, ... 
and by performing measurement only if $n+m =  n_{vol}$.  
We fitted the double-Gaussian  (\ref{SkinB}) to the measured  
cross-diagonals and extracted the bifurcation shift $c[n_{vol}]$. 
The parametrization (\ref{csexp}) seems to 
reproduce our data best, and we used it to calculate the values of the 
bifurcation point $s^b$ and the slope $c_0$. 
We present these results in Fig. \ref{FigC0Sb}. 
As expected the bifurcation slope $c_0$ gradually grows when 
$\Delta$ is decreased. This results in the observed contraction of
the time-extent of the blob. The gradual disappearance of the
double peak structure with increasing $\Del$ might also apply to 
other functions of the triangulations, and that might explain why for 
instance the 'traditional' order parameter used to identify the
'B'-'C' transition is seemingly  insensitive to the new 'bifurcation'
transition we observe around $\Del =$0.3-0.4 for $K_0=2.2$.

\begin{figure}[h!]
\centering
\scalebox{0.53}{\includegraphics{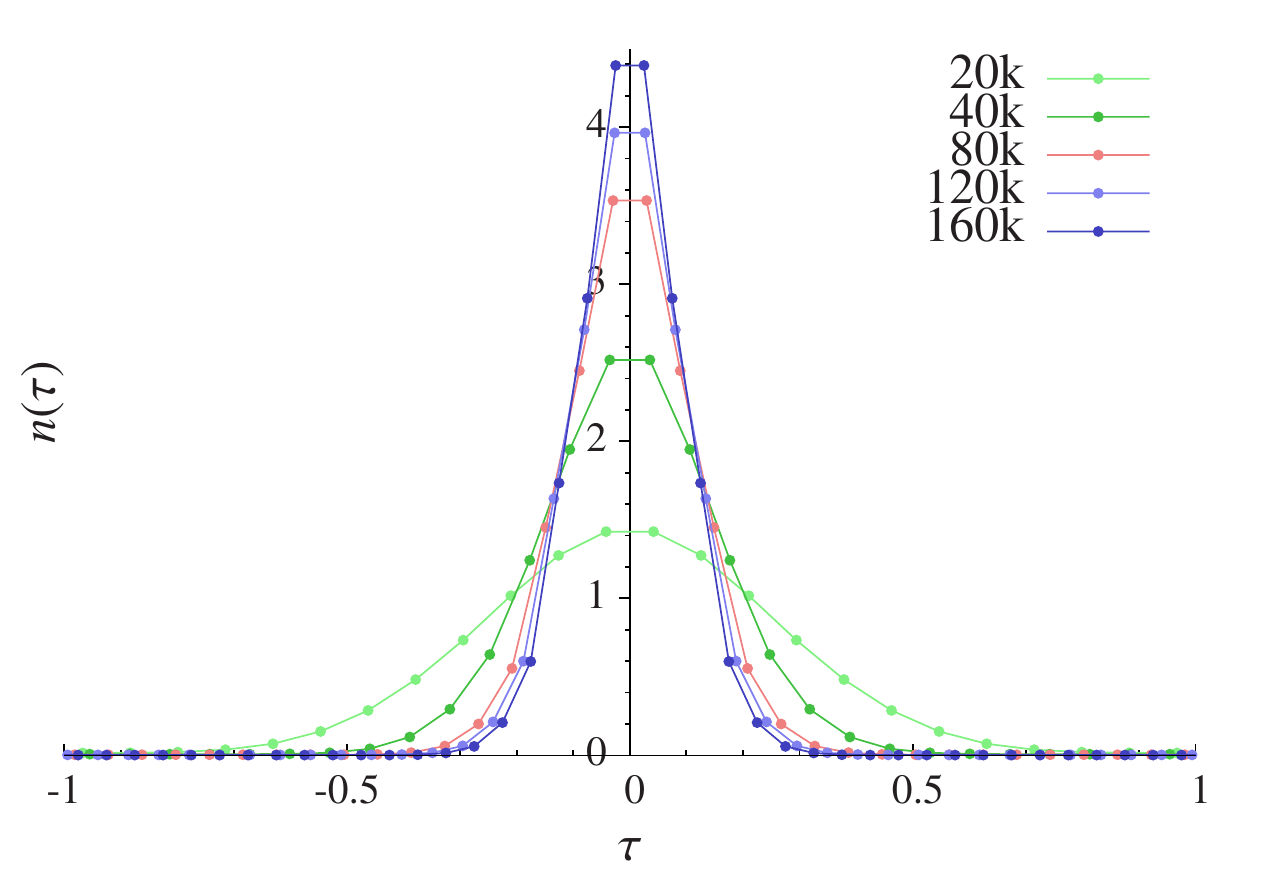}}
\scalebox{0.53}{\includegraphics{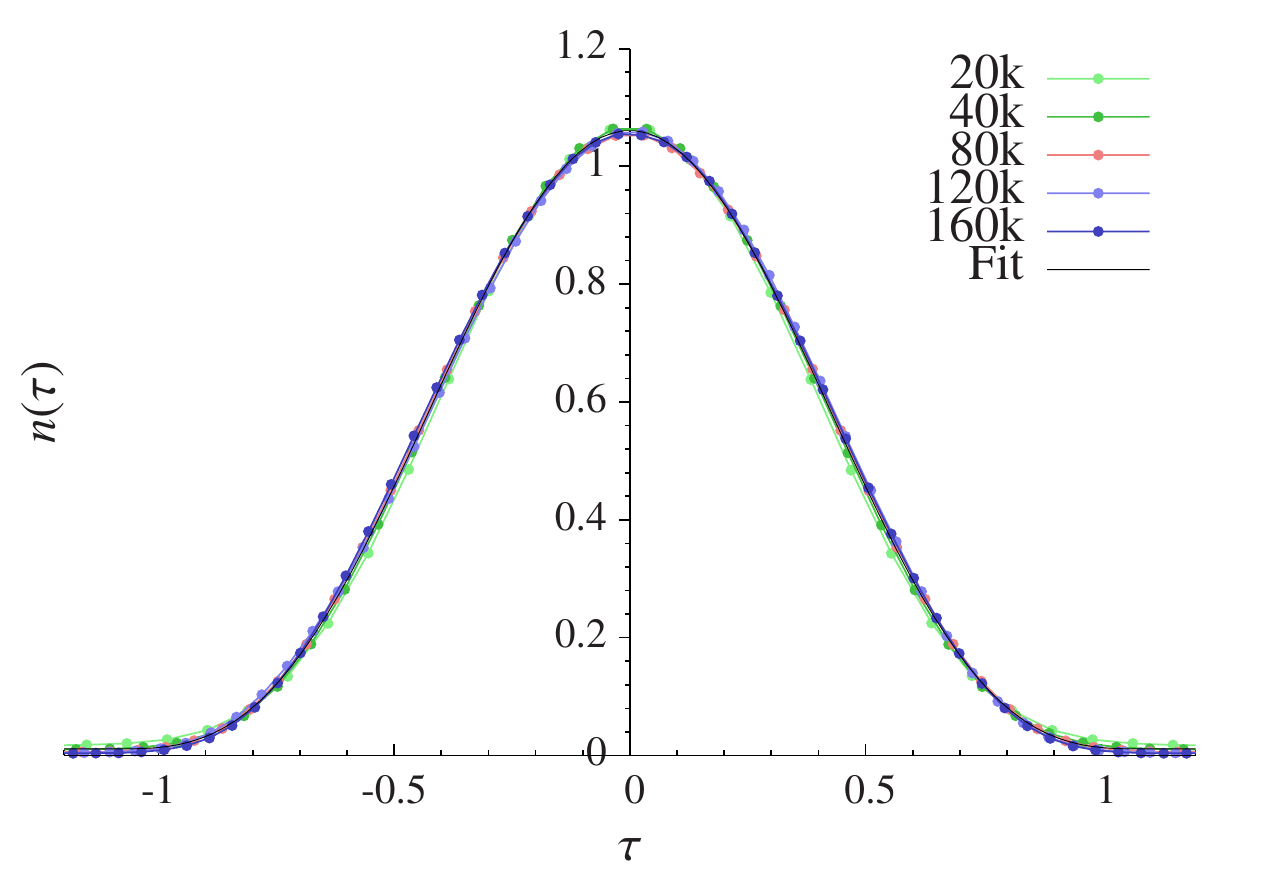}}
\caption{The  CDT  distributions of spatial volumes $ n_t$  
measured for different total four-volumes $\bar{N}_{41}$. The left figure 
shows the distributions in the 'bifurcation' phase 
($K_0=2.2$, $\Delta=0.125$), and the right figure the distributions
in the Sitter phase 'C'   ($K_0=2.2$, $\Delta=0.6$). 
Different colours correspond to different total volumes $\bar N_{41}$.  
The data were rescaled, according to: $\tau = t / \bar N_{41}^{1/d_H}$ and
$n(\tau) = n_t  / \bar N_{41}^{1-1/d_H}$,
to fit a single curve,  assuming the Hausdorff dimension $d_H= 4$ . 
The lack of scaling in the new phase is an important 
difference compared to the generic phase 'C'. }
\label{Fig13}
\end{figure}

\begin{figure}[h!]
\centering
\scalebox{0.53}{\includegraphics{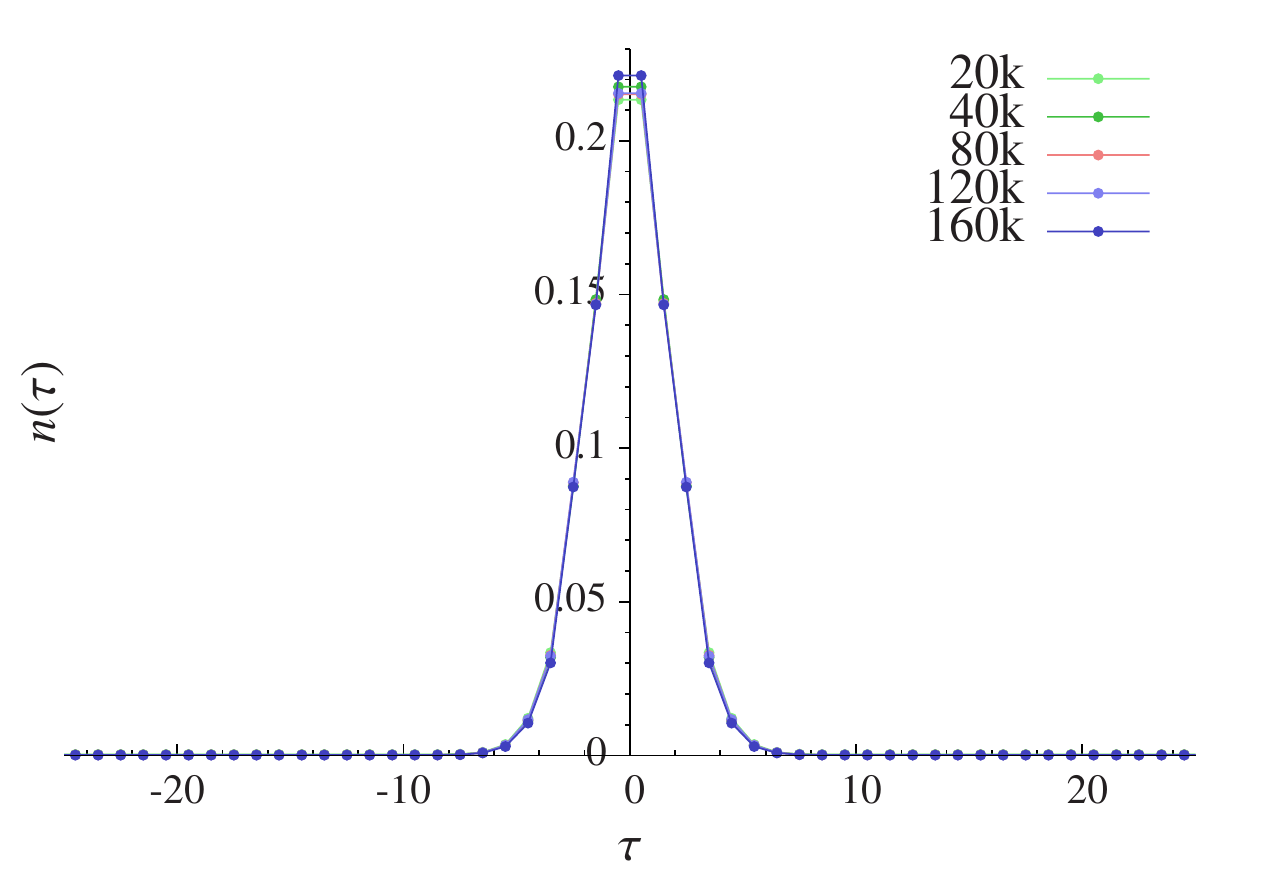}}
\scalebox{0.53}{\includegraphics{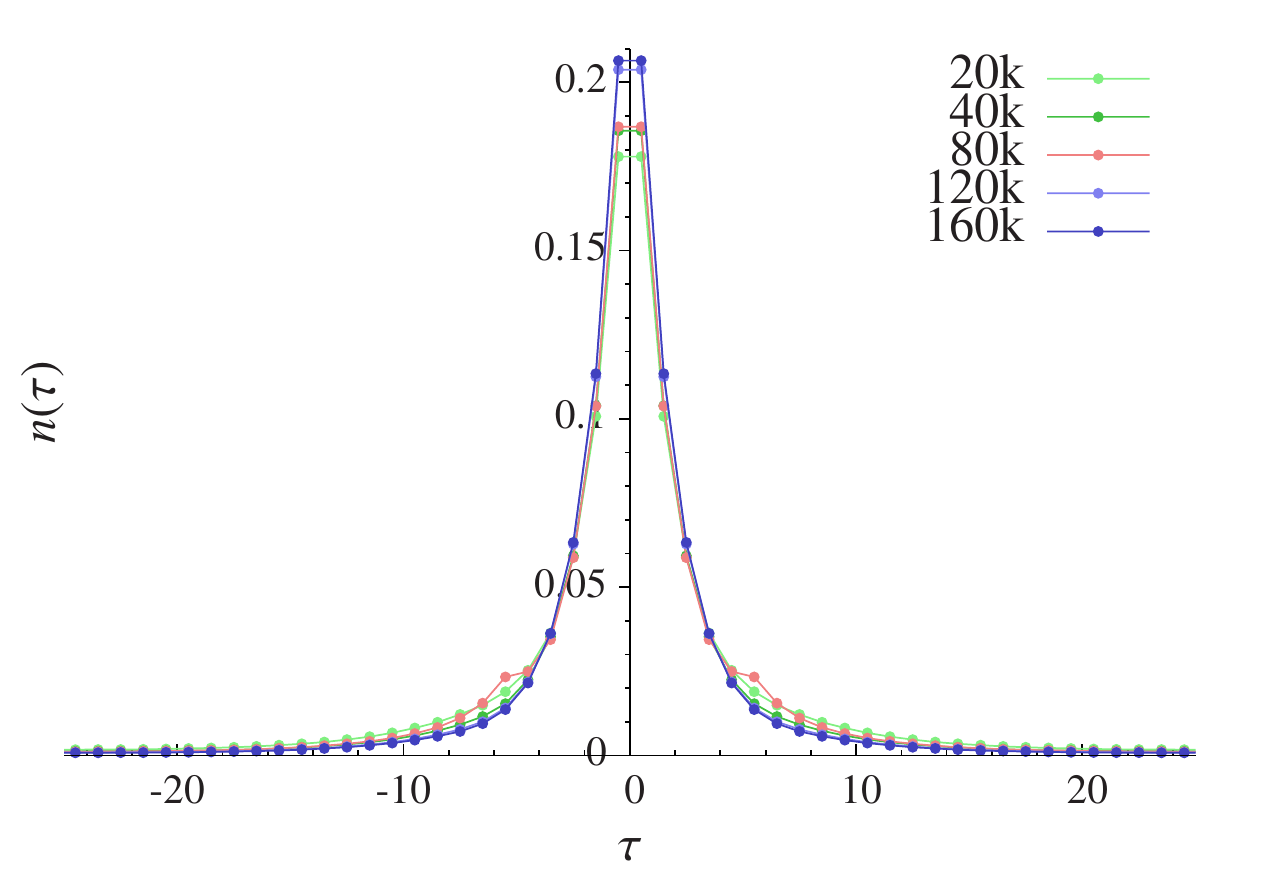}}
\caption{The left figure shows   CDT  distributions of spatial 
volumes $n_t$ measured  
in the  'bifurcation' phase ($K_0=2.2$, $\Delta=0.125$). The right figure
shows the $n_t$ distributions generated from 
the 'effective' Monte Carlo model (\ref{bifmodel}) with 
bifurcation slope $c_0=0.3$. Different colours correspond 
to different total volumes $\bar N_{41}$. 
The data were rescaled, according to: $\tau = t / \bar N_{41}^{1/d_H}$ and
$n(\tau) = n_t  / \bar N_{41}^{1-1/d_H}$, to fit a single 
curve, assuming the  Hausdorff dimension $d_H=\infty$.}
\label{bifscaling}
\end{figure}

Summarizing, we conclude that the new 
{\it 'bifurcation' phase} may exist in four-dimensional CDT. 
This phase should lie between the 'B' and 'C' phases. Its generic 
3-volume (temporal) distribution measured in 'full-CDT' with large $t_{tot}$ 
has  blob structures resembling those found in  phase 'C', but  
 there seem to be  important differences.
As an example we present the spatial volume distributions  
for different total volumes ($\bar N_{41}$) measured  inside the new phase 
for $K_0=2.2$ and $\Delta=0.125$  (Fig. \ref{Fig13}). The volume 
distributions were rescaled both in time ($\tau = t / \bar N_{41}^{1/d_H}$) 
and space  ($n(\tau) = n_t  / \bar N_{41}^{1-1/d_H}$) with $d_H=4$.  
In generic phase 'C' ($\Delta=0.6$) the Hausdorff dimension $d_H=4$, 
and the rescaled volume distributions fall onto a universal curve. 
The scaling for $\Delta=0.125$ is evidently  different. This is a 
strong argument in favour of the existence  of a genuinely  new phase. 
In fact the scaling (or rather lack of scaling) 
of volume profiles in the bifurcation 
region indicates that $d_H\to \infty$ for large volumes. This is  
consistent with our effective transfer matrix model 
(\ref{bifmodel}) and is characteristic of the generic phase 'B' - 
see Fig. \ref{bifscaling}. However, there is also no reason 
to doubt that the 'traditional' 'B'-'C' transition is still there, 
so seemingly we have discovered a new phase 
separating the 'old' phase 'C' and  phase 'B'.

\section{Summary and conclusions}

The recently introduced effective transfer matrix  labeled only 
by the spatial volume \cite{TM1} provides an interesting  tool for analyzing 
Causal Dynamical Triangulations in four dimensions.
Using Monte Carlo simulations of the complete CDT theory 
and the factorization (\ref{Pn1nttot}), 
we have determined the effective transfer matrix.

Assuming validity of decomposition (\ref{Pn1nttot}) 
we introduced {\it an effective transfer matrix model} reducing 
the degrees of freedom only to spatial volumes, 
i.e. a model which neglects completely the internal structure of slices.
The simplified model reproduces perfectly all results obtained so far 
using the full model of four-dimensional  CDT.

In phase 'C' the effective  Lagrangian
\beq\label{janx1} 
L_{C}(n,m)
=\frac{1}{\Gamma} \left[ 
\frac{(n - m)^2}{n + m - 2 n_0} + \mu \left( \frac{n + m}{2}\right)^{1/3} - \lambda \left( \frac{n + m}{2}\right) - \delta \left( \frac{n + m}{2}\right)^{- \rho}  \right] 
\eeq
describes very well the measured transfer matrix  
and can  seemingly be used also for large volumes.

Our present investigations were partly inspired by a recent 
study \cite{burda} which used  the effective Lagrangian 
\beq\label{janx2}
L_{eff} = c_1 \frac{2(n-m)^2}{ n+m} +c_2 \frac{m^{1/3}+n^{1/3}}{2} 
\eeq
to define an effective one-dimensional transfer matrix. The authors 
studied the phase diagram as a function of the coupling
constants $c_1$ and $c_2$. Interestingly, they observed a phase diagram 
which was quite similar to the CDT phase diagram. Clearly the effective 
Lagrangian \rf{janx2}  resembles the more precisely determined
Lagrangian \rf{janx1} which can be used to represent the full 
CDT model in phase 'C'. Thus we wanted to 
check, using the full CDT model, if \rf{janx2} could really be viewed
as giving a precise description of the full CDT model in phase 
'A' and 'B' for certain choices of $c_1$ and $c_2$. It turned out that this was
not really  the case.     

We provide strong evidence that inside the 'uncorrelated' phase 'A'  
the effective Lagrangian takes the form
$$
L_A(n,m)= \mu \left(n^{ \alpha} + m^{ \alpha} \right) - \lambda(n+m) \ ,
$$
with $\alpha \neq 1/3$. The absence of a kinetic term can be interpreted 
as a causal disconnection of different time slices, i.e. 
the phenomenon of 'asymptotic silence' observed both in classical and 
quantum  approaches to gravity in the regime of extreme 
curvatures/energy densities \cite{mielczarek}. In this context 
the 'uncorrelated' phase might gain some physical meaning.

The situation is more difficult in the 'collapsed' phase 'B'. 
Inside this phase the transfer matrix can be parametrized as follows:
$$
\braket{n|M_B|m} =
{\cal N}[n+m]
\left[ 
\exp \left( -\frac{\Big((m-n) -\big[ c_0(n+m-s^b) \big]_+\Big)^2}{\Gamma(n+m-2 n_0)}\right) + \right.
$$
$$
\left.
+\exp \left( -\frac{\Big((m-n) +\big[ c_0(n+m-s^b) \big]_+\Big)^2}{\Gamma(n+m-2 n_0)}\right) 
\right] \ ,
$$
where: $[.]_+ = \max(. , 0)$.
The properties of the spatial volume distribution depend strongly 
on the parameter $c_0$, while a new kind of  phase transition seems
to be related with the $s^b \to \infty$ limit.  
We showed by direct measurement that this kind of  phase transition 
occurs for the $\Delta$ coupling constant considerably larger than the $\Del$
value where the 'traditional' 'B'-'C' phase transition found in previous 
studies occurs. It points 
to a new 'bifurcation' phase separating the 'B' and 'C' phases. 
That such a putative phase is not an artifact of the effective 
transfer matrix model is supported by measurements performed 
in the full CDT theory. They show that the scaling of the 
spatial volume distribution $n_t$  as a function of the total 
four-volume $\bar{N}_{41}$ changes when we get close to the  
'old' 'B'-'C' transition. Deep in 
phase 'C' the scaling is canonical, i.e. $n_t \propto \bar{N}_{41}^{3/4}$
for spatial slices in the 'blob', but in the new phase 
the scaling seems to be closer to $n_t \propto \bar{N}_{41}$.
Such a different scaling of the spatial volume distribution and thus 
also of the time extent of the ``blob'' constituting the 
``physical'' universe  is interesting. It indicates
that the 'B'-'C' phase transition line might be associated with 
an asymmetric scaling of space and time, precisely as is 
assumed in Ho\v{r}ava-Lifshitz gravity \cite{HL}. The new 
'bifurcation' phase could be a genuinely new phase separating 
the 'B' phase from the ``ordinary'' 'C' phase.
The exact nature of this new phase and its transition regions
deserve further studies.

\section*{Acknowledgments}
JG-S acknowledges the Polish National
Science Centre (NCN) support via the grant
DEC-2012/05/N/ST2/02698. 
JA and AG were supported by the ERC-Advance grant 291092,
``Exploring the Quantum Universe'' (EQU). JA acknowledges support 
of FNU, the Free Danish Research Council, from the grant 
``quantum gravity and the role of black holes''. JA was supported 
in part by Perimeter Institute of Theoretical Physics.
Research at Perimeter Institute is supported by the Government of Canada
through Industry Canada and by the Province of Ontario through the 
Ministry of Economic Development \& Innovation.

\noindent The authors thank Daniel Coumbe for careful proofreading and fruitful discussions.

\end{document}